\documentclass[twocolumn, superscriptaddress]{revtex4-1}

\usepackage{hyperref}
\usepackage{amsmath,amssymb}
\usepackage{graphicx}
\usepackage{multirow}
\usepackage{blindtext}
\usepackage{color}

\usepackage{natbib}

\def \bsym {\boldsymbol}
\def \n {\nonumber\\}

\begin{document}

\title{Quantum anomalous Hall effect and electric-field-induced topological phase
transition in AB-stacked MoTe${}_2$/WSe${}_2$ moir\'e heterobilayers}

\author{Yao-Wen Chang}
\affiliation{Physics Division, National Center for Theoretical Sciences, Taipei 10617,
Taiwan}
\affiliation{Research Center for Applied Sciences, Academia Sinica, Taipei 11529, Taiwan}

\author{Yia-Chung Chang}
\affiliation{Research Center for Applied Sciences, Academia Sinica, Taipei 11529, Taiwan}
\begin{abstract}
We propose a new mechanism to explain the quantum anomalous Hall (QAH) effect and the
electric-field-induced topological phase transition in AB-stacked MoTe${}_2$/WSe${}_2$
moir\'e heterobilayers at $\nu=1$ hole filling. We suggest that the Chern band of the QAH
state is generated from an intrinsic band inversion composed of the highest two moir\'e
hole bands with opposite valley numbers and a gap opening induced by two
Coulomb-interaction-driven magnetic orders. These magnetic orders, including an in-plane
$120^{\circ}$-N\'eel order and an in-plane ferromagnetic order, interact with moir\'e
bands via corresponding in-plane exchange fields. The N\'eel order ensures the insulating
gap, the ferromagnetic order induces the non-zero Chern number, and both orders contribute
to time-reversal symmetry (TRS) breaking. The N\'eel order is acquired from the
Hartree-Fock exchange interaction, and the formation of ferromagnetic order is attributed
to interlayer-exciton condensation and exciton ferromagnetism. The exciton ferromagnetism
can be demonstrated by excitonic Bose-Hubbard physics and Berezinskii-Kosterlitz-Thouless
(BKT) transition. In low electric fields, the equilibrium state is a Mott-insulator state.
At a certain electric field, a correlated insulating state composed of the hole-occupied
band and the exciton condensate becomes a new thermodynamically stable phase, and the
topological phase transition occurs as the ferromagnetic order emerges. The consistency
between the present theory and experimental observations is discussed. Experimental
observations, including the spin-polarized/valley-coherent nature of the QAH state, the
absence of charge gap closure at the topological phase transition, the canted spin
texture, and the insulator-to-metal transition, are interpreted by the mechanism.
\end{abstract}

\maketitle

\section{Introduction}

Recently, quantum anomalous Hall (QAH) insulators and related topological materials have
drawn a lot of attention from scientists for their fundamental importance and potential to
design quantum devices\cite{konig2008quantum, bernevig2013topological,
hohenadler2013correlation, weng2015quantum, liu2016quantum}. A QAH (insulating) state is a
two-dimensional insulator that carries a chiral edge state exhibiting a quantized Hall
conductance in the unit of $e^2/h$, and the quantum Hall conductance along with zero
longitudinal resistance in the absence of an external magnetic field is known as the QAH
effect\cite{bernevig2013topological, hohenadler2013correlation, weng2015quantum,
liu2016quantum}. Moir\'e material is a new platform for studying the QAH
effect\cite{chen2020tunable, serlin2020intrinsic, tschirhart2021imaging, zhang2019nearly,
wu2019topological}. Recently, a QAH state in AB-stacked MoTe${}_2$/WSe${}_2$ moir\'e
heterobilayers and an electric-field-induced topological phase transition were observed at
$\nu=1$ hole filling under an out-of-plane electric field\cite{exp0, exp1, exp2}. Some
experimental observations, such as the spin-polarized/valley-coherent nature of the QAH
state\cite{exp1} and the absence of charge gap closure at the topological phase
transition\cite{exp0}, are unique among related materials. Several theories have been
proposed to explain the mechanism and observations\cite{theory1, theory2, theory3,
theory4, theory5, theory6, theory7, theory8, theory9}, but some questions remain.

A suitable theory to explain the QAH effect in AB-stacked MoTe${}_2$/WSe${}_2$
heterobilayers should meet certain theoretical criteria and be able to explain related
experimental observations. Theoretically, for an insulator exhibiting the QAH effect, it
must contain at least an occupied band that is topological non-trivial and carries a
non-zero Chern number (i.e. a Chern band), and time-reversal symmetry (TRS) of the
insulator must be broken\cite{bernevig2013topological, hohenadler2013correlation,
weng2015quantum, liu2016quantum}. Experimentally, in addition to the QAH effect,
AB-stacked MoTe${}_2$/WSe${}_2$ heterobilayers also show the following properties:
\begin{enumerate}
\item At a small electric field and $\nu=1$ hole filling, the longitudinal resistance
diverges rapidly as temperature decreases, indicating a Mott-insulator state\cite{exp0}.
\item The MoTe${}_2$ valence band maximum is about $300$ meV above the WSe${}_2$ valence
band maximum in the absence of an electric field. A topological phase transition between a
topological trivial insulating state and the QAH state occurs at $\nu=1$ hole filling as
the electric field contributes about $-172$ meV shift (with $0.66$ V/nm electric-field
strength and $2.6$ e$\cdot$\AA\; interlayer dipole moment\cite{li2021continuous}) to the
valence-band energy offset. However, no charge gap closure is found at the
transition\cite{exp0}.
\item The magnetic-field dependence of transverse resistances of the QAH state was
studied. A magnetic hysteresis with a sharp magnetic switching in low temperature was
observed, and the onset of magnetic ordering is at approximately $5\sim 6$ K\cite{exp0}.
\item The charge gap of the insulating state decreases continuously as the electric field
increases. An insulator-to-metal transition occurs as the electric-field strength reaches
$0.70$ V/nm. The metallic state seems to follow a Fermi liquid behavior at low
temperatures and converges to a finite resistance in the zero-temperature
limit\cite{exp0}.
\item The relative alignment of the spontaneous spin (valley) polarization in the moir\'e
heterobilayer in the QAH state was studied by the magnetic circular dichroism (MCD) of the
attractive polaron feature in each layer. It was found that the QAH ground state is
consistent with a spin-polarized/valley-coherent state across two layers, in which the
spin polarization is aligned\cite{exp1}.
\item The magnetic-field dependence of the out-of-plane spin polarization was also studied
by the MCD. The maximum MCD signals in both layers increase monotonically with increasing
magnetic fields until saturating, as the transverse resistance is quantized near zero
magnetic field and does not depend on the magnetic field. It implies that full spin
polarization is not necessary for quantized Hall transport, and a canted spin texture
could exist\cite{exp1}.
\item Evidence of quantum spin Hall (QSH) effect was observed at $\nu=2$ hole filling. A
band-to-QSH insulator transition occurs at $-130$ meV energy shift ($0.50$ V/nm
electric-field strength), and charge gap closure and reopening were found\cite{exp0,exp2}.
\end{enumerate}
In addition to these experimental observations in AB-stacked MoTe${}_2$/WSe${}_2$
heterobilayers, a continuous Mott transition is observed in AA-stacked
MoTe${}_2$/WSe${}_2$ heterobilayers at $\nu=1$ hole filling, but no QAH effect or QSH
effect is found\cite{li2021continuous}. Some theoretical works have explained some of
these properties, but a theory consistent with all experimental observations is still
elusive.

A critical issue of the QAH effect in the present system is the mechanism of
electric-field-induced topological phase transition at $\nu=1$ filling. The topological
phase transition bridges a topological trivial insulating state and the QAH insulating
state. Theoretical works to study the QAH effect are supposed to propose a mechanism to
explain the transition. In Ref.~\cite{theory2,theory4,theory5,theory6}, the topological
phase transition is explained by a band-inversion mechanism involving Coulomb interaction
and interlayer tunneling. Based on this mechanism, a band inversion between the moir\'e
bands at the MoTe${}_2$ layer and the WSe${}_2$ layer occurs due to electric-field-induced
band-energy shift. A topological gap is opened by interlayer tunneling. The TRS is broken
by a Coulomb-interaction-driven valley polarization of holes. A hole-occupied and
valley-polarized Chern band is formed, and the topological phase transition occurs. In
Ref.~\cite{theory7,theory8,theory9}, the band inversion between the moir\'e bands at
different layers is also induced by the electric field, but the gap opening here is
induced by Coulomb-interaction-driven topological exciton condensation. The exciton
condensate is a spin-polarized/valley-coherent state, and it also contributes to the TRS
breaking. The topological gap is opened and a hole-occupied Chern band is formed via an
inter-band exchange interaction induced by the exciton condensate. In Ref.~\cite{theory1},
the Chern band is generated by geometry-relaxation-induced pseudo-magnetic field and
intrinsic band inversion. The TRS is broken and the topological phase transition occurs
due to the interaction-driven valley polarization. In Ref.~\cite{theory3}, the moir\'e
band structure is studied by the model of a Dirac hole quasiparticle in a moir\'e
potential. The Chern band is formed and the topological phase transition occurs due to an
electric-field-induced phase-angle tuning for the moir\'e potential. The TRS is also
broken by the interaction-driven valley polarization.

For mechanisms proposed by Ref.~\cite{theory2, theory4, theory5, theory6, theory7,
theory8, theory9}, the electric-field-induced band inversion involves a charge gap closure
and reopening\cite{weng2015quantum}, yet it is not consistent with the observation of the
absence of charge gap closure. For mechanisms proposed by Ref.~\cite{theory1,theory3}, the
electric-field-induced band inversion is no longer required, but the
spin-polarized/valley-coherent state observed in the QAH insulator is not explained.
Besides, there is still no widely accepted explanation for the insulator-to-metal
transition or the canted spin texture.

In this article, a new mechanism for explaining the QAH state and the topological phase
transition in AB-stacked MoTe${}_2$/WSe${}_2$ heterobilayers is proposed. The Chern band
is generated from an intrinsic band inversion and a gap opening. The intrinsic band
inversion is composed of the highest two moir\'e hole bands with opposite valley numbers.
The gap opening is induced by two Coulomb-interaction-driven magnetic orders, an in-plane
$120^{\circ}$-N\'eel order and an in-plane ferromagnetic order. The N\'eel order ensures
the insulating gap, the ferromagnetic order induces the non-zero Chern number, and both
orders contribute to TRS breaking. The N\'eel order is acquired from the Hartree-Fock
exchange interaction, and the ferromagnetic order is attributed to interlayer-exciton
condensation. In low electric fields, the equilibrium state is a Mott-insulator state. At
a certain electric field, a correlated insulating state composed of the hole-occupied band
and the exciton condensate becomes a new thermodynamically stable phase, and the
topological phase transition occurs as the ferromagnetic order emerges. Since the band
inversion is intrinsic and the gap is opened before the topological phase transition,
there is no charge gap closure. The QAH state being spin-polarized/valley-coherent across
two layers is consistent with the interlayer exciton condensate. The insulator-to-metal
transition can be interpreted as an exciton Mott transition. The canted spin texture is
attributed to a coexistence of the N\'eel order and field-induced valley polarization of
holes. In Sec.~\ref{sec:moire_model}, the continuum model for the moir\'e heterobilayers
under magnetic and exchange fields is introduced, and the symmetry is discussed. In
Sec.~\ref{sec:chern_band}, the origin of Chern bands in the moir\'e heterobilayers is
studied. Effects of in-plane N\'eel order, in-plane ferromagnetic order, and field-induced
valley polarization on moir\'e bands are discussed. In Sec.~\ref{sec:trs_breaking}, the
concepts of interlayer-exciton condensation, exciton ferromagnetism, and exciton Mott
transition are introduced. Finally, in Sec.~\ref{sec:discussion}, the consistency between
the theory and experimental observations is discussed. Derivations and formulations for
studying moir\'e band structures and exciton condensation are given in
Appendix~\ref{sec:moire_band} and Appendix~\ref{sec:exciton_condensation}.

\section{Moir\'e superlattice\label{sec:moire_model}}

In this section, the geometry, model, symmetry, and band structure of AB-stacked
MoTe${}_2$/WSe${}_2$ heterobilayers are introduced. In Sec.~\ref{sec:geometry}, the
moir\'e superlattice and the moir\'e reciprocal lattice are introduced. In
Sec.~\ref{sec:model}, the continuum model of the moir\'e heterobilayer under an
out-of-plane magnetic field and in-plane exchange field is introduced. In
Sec.~\ref{sec:symmetry}, symmetry of the continuum model is discussed. Finally, in
Sec.~\ref{sec:bandstructure}, calculation of single-particle band structure by using
plane-wave method is performed, and the moir\'e band structures of AB-stacked
MoTe${}_2$/WSe${}_2$ heterobilayers are shown. The details of using the plane-wave method
and Hartree-Fock approximation to calculate moir\'e band structures are introduced in
Appendix~\ref{sec:moire_band}.

\subsection{Geometry\label{sec:geometry}}

\begin{figure}
  \includegraphics[width=0.95\linewidth]{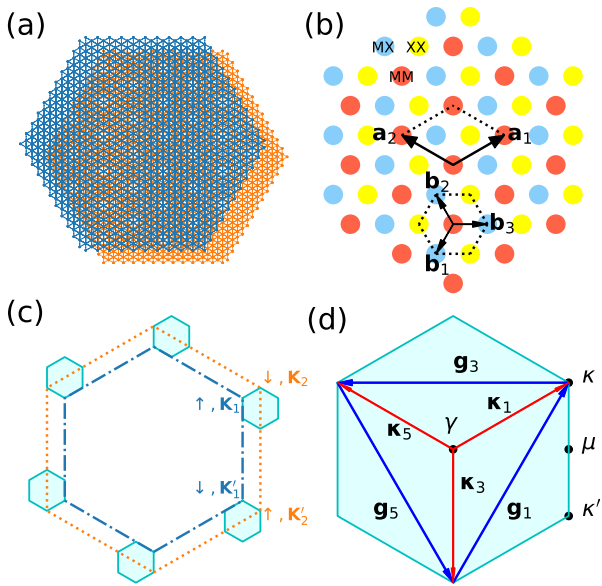}
  \caption{(a) Schematic plot (the ratio of lattice constants is not accurate) of the
  moir\'e heterobilayers. (b) The high-symmetry sites in the moir\'e superlattice. (c)
  Schematic plot of the Brillouin zone of a MoTe${}_2$ monolayer (inside blue dash-dot
  hexagonal) and the Brillouin zone of a WSe${}_2$ monolayer (inside orange dot
  hexagonal). The filled zone is the MBZ of AB-stacked MoTe${}_2$/WSe${}_2$
  heterobilayers. (d) A closer look at the MBZ and high-symmetry points.}
  \label{fig:moire_BZ}
\end{figure}

A moir\'e superlattice is formed due to the mismatch between the MoTe${}_2$ hexagonal
lattice and the WSe${}_2$ hexagonal lattice with different lattice constants. Schematic
plots of the moir\'e superlattice are illustrated in Fig.~\ref{fig:moire_BZ} (a), (b). The
lattice constant of the moir\'e superlattice $a_{\text{M}}$ as a function of the lattice
mismatch $\delta=|a'_0-a_0|/a_0$ is given by $a_{\text{M}}={(1+\delta)a_0}/\delta$, with
$a_0$, $a'_0$ the lattice constants for the atomistic lattices. The moir\'e superlattice
can be seen as a triangular lattice with local geometry in the unit cell. The periodicity
of a triangular lattice can be studied by primitive vectors, and the local geometry can be
indicated by basis vectors. The primitive vectors of the moir\'e superlattice can be
defined as $\bsym{a}_1 =a_{\text{M}}[(\sqrt{3}/2)\mathbf{e}_{x}+(1/2)\mathbf{e}_{y}]$,
$\bsym{a}_2=a_{\text{M}}[(-\sqrt{3}/2)\mathbf{e}_{x}+(1/2)\mathbf{e}_{y}]$. The basis
vectors of the moir\'e superlattice are given by
$\bsym{b}_1=-\left(2\bsym{a}_{1}+\bsym{a}_{2}\right)/3$,
$\bsym{b}_2=\left(\bsym{a}_{1}+2\bsym{a}_{2}\right)/3$, and
$\bsym{b}_3=\left(\bsym{a}_{1}-\bsym{a}_{2}\right)/3$. In Fig.~\ref{fig:moire_BZ} (b),
these primitive and basis vectors are shown.

The moir\'e reciprocal lattice and moir\'e Brillouin zone (MBZ) can also be utilized to
demonstrate the geometry of the moir\'e superlattice. The MBZ of the moir\'e superlattice
is illustrated in Fig.~\ref{fig:moire_BZ} (c), (d). As can be seen in
Fig.~\ref{fig:moire_BZ} (c), the MBZ can be constructed by the geometry difference between
the Brillouin zone of a MoTe${}_2$ monolayer and the Brillouin zone of a WSe${}_2$
monolayer. The moir\'e reciprocal lattice is the periodic repeat of the Brillouin zone.
For the moir\'e reciprocal lattice, the reciprocal primitive vectors can be defined by
$\bsym{g}_{i}\cdot\bsym{a}_{j}=2\pi\delta_{ij}$ for $i,j=1,2$. We get
$\bsym{g}_{1}=\sqrt{3}k_{\text{M}}[(1/2)\mathbf{e}_{x} +(\sqrt{3}/2)\mathbf{e}_{y}]$,
$\bsym{g}_{2} =\sqrt{3}k_{\text{M}}[(-1/2)\mathbf{e}_{x} +(\sqrt{3}/2)\mathbf{e}_{y}]$,
with $k_{\text{M}} =4\pi/(3a_{\text{M}})$. A set of reciprocal primitive vectors can be
defined as
\begin{eqnarray}
  \bsym{g}_{j}
  =
  \sqrt{3}k_{\text{M}}\left[\mathbf{e}_{x}\cos({j\pi}/{3})
  +\mathbf{e}_{y}\sin({j\pi}/{3})\right],
\end{eqnarray}
with $j=1,2,\cdots,6$. The high-symmetry points are indicated in Fig.~\ref{fig:moire_BZ}
(d). The vector connects between $\kappa$ and $\gamma$ is given by
$\bsym{\kappa}_1=\left(2\bsym{g}_{1}-\bsym{g}_{2}\right)/3$ and the vector connects
between $\kappa'$ and $\gamma$ is given by
$\bsym{\kappa}_2=\left(\bsym{g}_{1}-2\bsym{g}_{2}\right)/3$. These two vectors
$\bsym{\kappa}_1$ and $\bsym{\kappa}_2$ can be assigned as reciprocal basis vectors. A set
of reciprocal basis vectors is defined as
\begin{eqnarray}
  \bsym{\kappa}_{j}
  =
  k_{\text{M}}\left[\mathbf{e}_{x}\cos(\pi/6-{j\pi}/{3})
  +\mathbf{e}_{y}\sin(\pi/6-{j\pi}/{3})\right],
\end{eqnarray}
with $j=1,2,\cdots,6$. Part of the reciprocal primitive and basis vectors are shown in
Fig.~\ref{fig:moire_BZ} (d).

\subsection{Continuum model\label{sec:model}}

By knowing the geometry of the moir\'e superlattice and the MBZ, we can write down the
continuum model with effective-mass approximation. The continuum Hamiltonian for a hole in
the moir\'e heterobilayer is written as\cite{wu2019topological, wu2018hubbard}
\begin{eqnarray}
  H(\mathbf{r})
  =
  \begin{pmatrix}
  \underline{h}_{+}(\mathbf{r}) & \underline{\gamma}(\mathbf{r})\\
  \underline{\gamma}^{\dagger}(\mathbf{r}) & \underline{h}_{-}(\mathbf{r})
  \end{pmatrix},
  \label{eff_mass_hamil}
\end{eqnarray}
with
\begin{eqnarray}
\underline{h}_{\tau}(\mathbf{r})
=
\begin{pmatrix}
h_{\tau 1}(\mathbf{r}) & t_{\tau}(\mathbf{r})\\
t^*_{\tau}(\mathbf{r}) & h_{\tau 2}(\mathbf{r})
\end{pmatrix},\hskip1ex
\underline{\gamma}(\mathbf{r})
=
\begin{pmatrix}
\gamma_{1}(\mathbf{r}) & 0\\
0 & \gamma_{2}(\mathbf{r})
\end{pmatrix},\hskip1ex
\end{eqnarray}
where $h_{\tau l}(\mathbf{r})$ is the layer Hamiltonian with $l=1,2$ indicating the top,
bottom layers, $t_{\tau}(\mathbf{r})$ is the interlayer tunneling, and
$\gamma_{l}(\mathbf{r})$ is the in-plane exchange field. The interlayer tunneling is given
by
\begin{eqnarray}
  t_{\tau}(\mathbf{r})
  =
  w\left(1
  +e^{\mathtt{i}\tau\bsym{g}_1\cdot\mathbf{r}}
  +e^{\mathtt{i}\tau\bsym{g}_2\cdot\mathbf{r}}\right),
\end{eqnarray}
where $w$ is the interlayer-tunneling coupling. The layer Hamiltonian is given by
\begin{eqnarray}
  h_{\tau l}(\mathbf{r})
  &=&
  \epsilon_{\tau l}+\frac{|\mathbf{p}-\tau\bsym{\kappa}_{l}|^2}{2m_{l}}-V_{l}(\mathbf{r}),
\end{eqnarray}
where $\epsilon_{\tau l}$ is the band-edge energy and $V_{l}(\mathbf{r})$ is the moir\'e
potential. The band-edge energy is given by
\begin{eqnarray}
  \epsilon_{\tau l}
  =
  [D+(-1)^{l}D]/2
  -(sg_{\text{spin}}+\tau g_{\text{valley}})\mu_{\text{B}}{B}_z,
  \label{band_edge1}
\end{eqnarray}
with $D$ the valence-band energy offset, ${B}_{z}$ the external out-of-plane magnetic
field, $\mu_{\text{B}}$ the Bohr magneton, $s=+,-$ the direction for $\uparrow,\downarrow$
spin, $g_{\text{spin}}$ the spin g-factor and $g_{\text{valley}}$ the valley g-factor.
Since $g_{\text{spin}}\gg g_{\text{valley}}$, accordingly, $g_{\text{valley}}\simeq{0}$ is
assumed. Based on Fig.~\ref{fig:moire_BZ} (c), the spin directions of $\mathbf{K}_{l}$,
$\mathbf{K}'_{l}$ valleys at $l$-th layer are given by $s=-(-1)^{l}\tau$, with $\tau=+$
for $\mathbf{K}_{l}$ valley and $\tau=-$ for $\mathbf{K}'_{l}$ valley. The band-edge
energy can be rewritten as
\begin{eqnarray}
  \epsilon_{\tau 1}=-\tau M_z,\hskip2ex
  \epsilon_{\tau 2}={D}+\tau M_z
  \label{band_edge2}
\end{eqnarray}
with $M_z=g_{\text{spin}}\mu_{\text{B}}B_{z}$ an out-of-plane field-induced magnetization.
The moir\'e potential is given by
\begin{eqnarray}
  V_{l}(\mathbf{r})
  =
  (-1)^{l}2V\sum_{j=1,3,5}\sin\left(\bsym{g}_{j}\cdot\mathbf{r}\right)
  \label{moire_potential}
\end{eqnarray}
where $V$ is the potential depth. For the interlayer tunneling and the moir\'e potential,
note that $t_{\tau}(\bsym{b}_1) =t_{\tau}(\bsym{b}_2) =t_{\tau}(\bsym{b}_3)=0$ and
$t_{\tau}(\mathbf{0}) =t_{\tau}(\bsym{a}_{1}) =t_{\tau}(\bsym{a}_{2})=3w$,
$V_{l}(\bsym{b}_1) =V_{l}(\bsym{b}_2) =V_{l}(\bsym{b}_3) =-V_{l}(-\bsym{b}_1)
=-V_{l}(-\bsym{b}_2) =-V_{l}(-\bsym{b}_3) =(-1)^{l}3\sqrt{3}V$ and $V_{l}(\mathbf{0})
=V_{l}(\bsym{a}_{1}) =V_{l}(\bsym{a}_{2})=0$. These points can be assigned at the
high-symmetry sites in the moir\'e superlattice. If the coordinate is transformed as
$\mathbf{r}\rightarrow\mathbf{r}+\bsym{b}_2$, the moir\'e potential and the interlayer
tunneling become
\begin{eqnarray}
  V_{l}(\mathbf{r}+\bsym{b}_2)
  &=&
  (-1)^{l}2V\sum_{j=1,3,5}\cos\left(\bsym{g}_{j}\cdot\mathbf{r}+{\pi}/{6}\right),
\end{eqnarray}
\begin{eqnarray}
  t_{\tau}(\mathbf{r}+\bsym{b}_2)
  &=&
  w\left[1+e^{\mathtt{i}\tau\left(\bsym{g}_1\cdot\mathbf{r}+{2\pi}/{3}\right)}
  +e^{\mathtt{i}\tau\left(\bsym{g}_2\cdot\mathbf{r}+{4\pi}/{3}\right)}\right].\hskip3ex
\end{eqnarray}
The later formulation of moir\'e potential and interlayer tunneling is more frequently
seen in literature, but the two formulations are equivalent.

The in-plane exchange field includes the contributions from a ferromagnetic exchange field
and a $120^{\circ}$-antiferromagnetic exchange field, which are generated from an in-plane
ferromagnetic order and an in-plane $120^{\circ}$-N\'eel order, respectively. The origins
of these two magnetic orders will be discussed in Sec.~\ref{sec:chern_band} and
Sec.~\ref{sec:trs_breaking}. Note that the in-plane exchange field is not uniformly
effective, since the magnetic order is originated from localized spins residing at each
moir\'e unit cell. The exchange field should show the same periodicity as the moir\'e
superlattice. The $120^{\circ}$-N\'eel order is a three-sublattice antiferromagnetic order
with the directions of spins at three sublattices being separated by $2\pi/3$ angular
difference\cite{leung1993spin, wu2018hubbard, pan2020band, zang2021hartree}. A general
form of the three-sublattice exchange field is written as
\begin{eqnarray}
  \gamma_{l}(\mathbf{r})
  &=&
  e^{\mathtt{i}2\bsym{\kappa}_l\cdot\mathbf{r}}
  \big[M_{A}\exp\left(\mathtt{i}\bsym{\kappa}_{1}\cdot(\mathbf{r}-\bsym{b}_1)\right)\n
  &&+M_{B}\exp\left(\mathtt{i}\bsym{\kappa}_{3}\cdot(\mathbf{r}-\bsym{b}_2)\right)\n
  &&+M_{C}\exp\left(\mathtt{i}\bsym{\kappa}_{5}\cdot(\mathbf{r}-\bsym{b}_3)\right)\big],
\end{eqnarray}
where $M_{A}$, $M_{B}$, $M_{C}$ are sublattice magnetizations. The phase term
$e^{\mathtt{i}2\bsym{\kappa}_l\cdot\mathbf{r}}$ is added to the exchange field to counter
the phase difference between single-particle states at two valleys. For the in-plane
ferromagnetic order, sublattice magnetizations follow the relation $M_{A}=M_{B}=M_{C}$.
The in-plane ferromagnetic exchange field is then given by
\begin{eqnarray}
  \gamma_{l}(\mathbf{r})
  &=&
  e^{\mathtt{i}2\bsym{\kappa}_l\cdot\mathbf{r}}M_{\perp}
  \sum_{j=1,3,5}\exp\left(\mathtt{i}\bsym{\kappa}_{j}\cdot\mathbf{r}\right),
\end{eqnarray}
with $M_{\perp}$ the in-plane ferromagnetic magnetization. On the other hand, for the
in-plane $120^{\circ}$-N\'eel order, sublattice magnetizations follow the relation
$M_{A}=\exp(\mathtt{i}2\pi/3)M_{B}=\exp(\mathtt{i}4\pi/3)M_{C}$. The in-plane
$120^{\circ}$-antiferromagnetic exchange field is given by
\begin{eqnarray}
  \gamma_{l}(\mathbf{r})
  &=&
  e^{\mathtt{i}2\bsym{\kappa}_l\cdot\mathbf{r}}M'_{\perp}
  \big[\exp\left(\mathtt{i}(\bsym{\kappa}_{1}\cdot\mathbf{r}+4\pi/3)\right)\n
  &&+\exp\left(\mathtt{i}(\bsym{\kappa}_{3}\cdot\mathbf{r}+2\pi/3)\right)
  +\exp\left(\mathtt{i}\bsym{\kappa}_{5}\cdot\mathbf{r}\right)\big],
\end{eqnarray}
with $M'_{\perp}$ the in-plane antiferromagnetic magnetization.

\subsection{Symmetry\label{sec:symmetry}}

The TRS and three-fold rotational ($C_3$) symmetry of the continuum model are discussed.
The time-reversal operation ($\Theta$) is defined by $\Theta
=\exp\left(-\mathtt{i}\pi\sigma_{y}/2\right)\mathcal{K}
=-\mathtt{i}\sigma_{y}\mathcal{K}$, where $\mathcal{K}$ is defined by
\begin{eqnarray}
  \mathcal{K}\mathtt{i}\mathcal{K}^{-1}=-\mathtt{i},\hskip2ex
  \mathcal{K}\mathbf{p}\mathcal{K}^{-1}=-\mathbf{p}.
\end{eqnarray}
The $-\mathtt{i}\sigma_{y}$ operation gives $-\mathtt{i}\sigma_{y}|+\rangle=|-\rangle$ and
$-\mathtt{i}\sigma_{y}|-\rangle=-|+\rangle$, where $|+\rangle$ and $|-\rangle$ are the
state kets with valley numbers $\tau=+,-$. It can be found that
$\Theta|\tau\rangle\left[h_{\tau l}(\mathbf{r})-\epsilon_{\tau
l}\right]\langle\tau|\Theta^{-1} = |-\tau\rangle\left[h_{-\tau
l}(\mathbf{r})-\epsilon_{-\tau l}\right]\langle-\tau|$, and
$\Theta|\tau\rangle{t}_{\tau}(\mathbf{r})\langle\tau|\Theta^{-1}
=|-\tau\rangle{t}_{-\tau}(\mathbf{r})\langle-\tau|$. For the out-of-plane magnetic field,
the time-reversal operation gives $\Theta|\tau\rangle\epsilon_{\tau
l}\langle\tau|\Theta^{-1}= |-\tau\rangle\epsilon_{\tau l}\langle-\tau|$. For the in-plane
exchange field, the time-reversal operation gives
$\Theta|+\rangle{\gamma}_{l}(\mathbf{r})\langle-|\Theta^{-1} =
-|-\rangle{\gamma}^*_{l}(\mathbf{r})\langle+|$ and
$\Theta|-\rangle{\gamma}^*_{l}(\mathbf{r})\langle+|\Theta^{-1}
=-|+\rangle{\gamma}_{l}(\mathbf{r})\langle-|$. Therefore, the continuum Hamiltonian in the
absence of a magnetic field and exchange field is invariant under the time-reversal
operation, and the directions of magnetizations are reversed $(M_z,M_{\perp},M'_{\perp})
\rightarrow (-M_z,-M_{\perp},-M'_{\perp})$ in the Hamiltonian with out-of-plane magnetic
fields and in-plane exchange fields. Therefore, the magnetizations
$M_z,M_{\perp},M'_{\perp}$ can be considered as TRS breaking terms.

To study the $C_3$ symmetry of the continuum model, it is convenient to apply a unitary
transformation to the Hamiltonian $\bar{H}(\mathbf{r})
=\mathcal{U}(\mathbf{r}){H}(\mathbf{r}) \mathcal{U}^{\dagger}(\mathbf{r})$, where the
unitary transformation matrix is given by
\begin{eqnarray}
  \mathcal{U}(\mathbf{r})
  &=&
  \begin{pmatrix}
  e^{-\mathtt{i}\bsym{\kappa}_{1}\cdot\mathbf{r}} & 0 & 0 & 0\\
  0 & e^{-\mathtt{i}\bsym{\kappa}_{2}\cdot\mathbf{r}} & 0 & 0\\
  0 & 0 & e^{\mathtt{i}\bsym{\kappa}_{1}\cdot\mathbf{r}} & 0\\
  0 & 0 & 0 & e^{\mathtt{i}\bsym{\kappa}_{2}\cdot\mathbf{r}}
  \end{pmatrix},
\end{eqnarray}
with the basis $(\tau,l)=(+,1),(+,2),(-,1),(-,2)$. The layer Hamiltonian becomes
\begin{eqnarray}
  \bar{h}_{\tau l}(\mathbf{r})
  &=&
  \epsilon_{\tau l}+\frac{|\mathbf{p}|^2}{2m_{l}}-V_{l}(\mathbf{r}),
\end{eqnarray}
and the interlayer tunneling becomes
\begin{eqnarray}
  \bar{t}_{\tau}(\mathbf{r})
  =
  w\sum_{j=1,3,5}\exp\left({\mathtt{i}\tau\bsym{\kappa}_j\cdot\mathbf{r}}\right).
\end{eqnarray}
The in-plane exchange field including both ferromagnetic magnetization and
antiferromagnetic magnetization is given by
\begin{eqnarray}
  \bar{\gamma}_{l}(\mathbf{r})
  &=&
  M_{\perp}\sum_{j=1,3,5}\exp\left(\mathtt{i}\bsym{\kappa}_{j}\cdot\mathbf{r}\right)\n
  &&+
  M'_{\perp}\sum_{j=1,3,5}
  \big[\exp\left(\mathtt{i}(\bsym{\kappa}_{1}\cdot\mathbf{r}+4\pi/3)\right)\n
  &&+\exp\left(\mathtt{i}(\bsym{\kappa}_{3}\cdot\mathbf{r}+2\pi/3)\right)
  +\exp\left(\mathtt{i}\bsym{\kappa}_{5}\cdot\mathbf{r}\right)\big].
\end{eqnarray}
The $C_3$ operation is defined by
\begin{eqnarray}
  C_3\bar{H}(\mathbf{r})C_{3}^{-1}
  =
  \bar{H}(\mathbb{R}_{[-2\pi/3]}\mathbf{r}),
\end{eqnarray}
where the rotational matrix satisfies
\begin{eqnarray}
  \mathbb{R}_{[\theta]}\bsym{v}=
  \begin{pmatrix}
  \cos\theta & \sin\theta\\
  -\sin\theta & \cos\theta
  \end{pmatrix}
  \begin{pmatrix}
  v_x\\ v_y
  \end{pmatrix}
\end{eqnarray}
for arbitrary vector $\bsym{v}$. The rotational matrix gives
$\mathbb{R}_{[4\pi/3]}\bsym{g}_1 =\mathbb{R}_{[2\pi/3]}\bsym{g}_3 =\bsym{g}_5$, and
$\mathbb{R}_{[4\pi/3]}\bsym{\kappa}_5 =\mathbb{R}_{[2\pi/3]}\bsym{\kappa}_3
=\bsym{\kappa}_1$. It is found that the $C_3$-symmetry is conserved for the continuum
model with the ferromagnetic exchange field. On the other hand, the in-plane
antiferromagnetic exchange field breaks the $C_3$ symmetry.

\subsection{Moir\'e band structure\label{sec:bandstructure}}

\begin{figure}
  \includegraphics[width=0.95\linewidth]{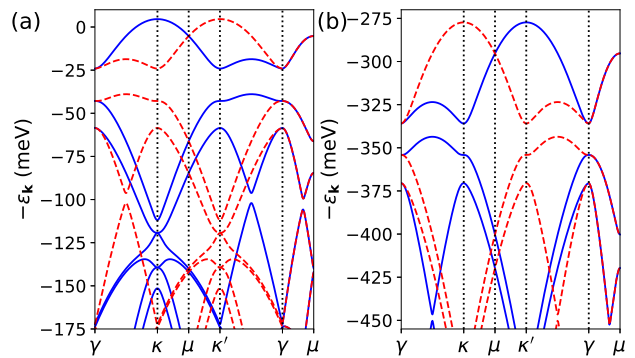}
  \caption{Moir\'e band structures of holes contributed from (a) the MoTe${}_2$ layer and
  (b) the WSe${}_2$ layer of a AB-stacked heterobilayer simulated by the continuum model
  and the plane-wave method in the absence of interlayer tunneling and magnetization. The
  blue solid line indicates $\tau=+$ and the red dash line indicates $\tau=-$. The high
  symmetry points are tagged in the MBZ in Fig.~\ref{fig:moire_BZ} (d).}
  \label{fig:moire_band}
\end{figure}

The effective hole mass of the MoTe${}_2$ is given by $m_{1}/m_0=0.62$ and the effective
hole mass of the WSe${}_2$ is $m_{2}/m_0=0.36$\cite{xiao2012coupled, kylanpaa2015binding}.
The parameters for the moir\'e heterobilayers are assumed to be $D=280$ meV, $V=10$ meV,
$a_{\text{M}}=50$ \AA, and $w=1.0$ meV. The interlayer-tunneling coupling is small because
the interlayer tunneling is spin-forbidden in the leading order approximation for the
moir\'e heterobilayers\cite{exp0, theory2}. The single-particle band structure of the
moir\'e superlattice, name as moir\'e band structure, can be calculated by using the
plane-wave basis function method (see appendix Sec.~\ref{sec:plane_wave}). While the
interlayer-tunneling coupling is small in comparing with the valence-band offset, the
moir\'e band structure is studied  in the absence of the interlayer tunneling, such that
the moir\'e bands can be assigned as contributions from different layers. In
Fig.~\ref{fig:moire_band}, moir\'e band structures for holes contributed from the
MoTe${}_2$ layer and the WSe${}_2$ layer of the moir\'e heterobilayers in the absence of
interlayer tunneling and magnetization are shown. Note that an intrinsic band inversion
locates across $\gamma-\mu$ lines in the MBZ between the highest two moir\'e hole bands
with opposite valley numbers in the MoTe${}_2$ layer. In Sec.~\ref{sec:chern_band}, we
will study how Coulomb interaction opens a gap at the intrinsically inverted moir\'e bands
and how the topological order emerges.

\section{interaction-driven Chern band\label{sec:chern_band}}

In this section, the formation of the Chern band in the QAH state is studied. The Chern
band is generated by opening a gap to break the intrinsic band inversion across
$\gamma-\mu$ lines. The gap is opened by in-plane exchange fields contributed from
corresponding Coulomb-interaction-driven in-plane $120^{\circ}$-N\'eel order and in-plane
ferromagnetic order. The N\'eel order ensures the insulating gap and the ferromagnetic
order generates the non-zero Chern number. In Sec.~\ref{sec:interacting}, the
many-particle Hamiltonian for Coulomb-interacting systems and Hartree-Fock approximation
are introduced. The method to calculate Chern numbers for single-particle band structures
in interacting systems is reviewed. In Sec.~\ref{sec:gap_opening}, the
$120^{\circ}$-N\'eel order is derived from the Hartree-Fock exchange interaction. The
competition between the $120^{\circ}$-N\'eel order and the valley polarization of holes
driven by an external magnetic field is studied. Effects of the N\'eel order and the
ferromagnetic order on the gap opening are discussed. In Sec.~\ref{sec:Chern_number},
Chern numbers of moir\'e bands are assigned by studying the winding numbers of Fock
pseudospin textures.

\subsection{Interacting systems\label{sec:interacting}}

To include the Coulomb interaction in the band-structure picture, we consider the
many-particle Hamiltonian for multi-component fermion fields
\begin{eqnarray}
  \hat{\mathcal{H}}
  &=&
  \sum_{ab}\int\hat{\Psi}^{\dagger}_{a}(\mathbf{r})H_{ab}(\mathbf{r})
  \hat{\Psi}_{b}(\mathbf{r})\text{d}^2r\n
  &&+\frac{1}{2}\sum_{ab}\int W_{ab}(\mathbf{r}_{12}):\hat{\rho}_{a}(\mathbf{r}_1)
  \hat{\rho}_{b}(\mathbf{r}_2):\text{d}^2r_1\text{d}^2r_2,
  \label{many_particle}
\end{eqnarray}
where $a=\{\tau,l\}$ is the component index. $H_{ab}(\mathbf{r})$ is the continuum
Hamiltonian given in Eq.~(\ref{eff_mass_hamil}), $W_{ab}(\mathbf{r}_{12})$ is the Coulomb
potential, $\hat{\Psi}^{\dagger}_{a}(\mathbf{r})$ and $\hat{\Psi}_{a}(\mathbf{r})$ are
fermion creation and annihilation operators for charges, and $\hat{\rho}_{a}(\mathbf{r})
=\hat{\Psi}^{\dagger}_{a}(\mathbf{r}) \hat{\Psi}_{a}(\mathbf{r})$ is the density operator.
The Chern number $\mathcal{C}$ of an insulator can be related to the quantized Hall
conductance by $\sigma_{\text{H}}=\frac{e^2}{2\pi}\mathcal{C}$. For interacting
many-particle systems, the Chern number contributed from the band structures can be
calculated by\cite{hohenadler2013correlation, so1985induced, ishikawa1987microscopic,
matsuyama1987quantization, wang2010topological}
\begin{eqnarray}
  \mathcal{C}
  &=&
  \frac{\epsilon_{\mu\nu\rho}}{6}\int
  \text{Tr}\left[
  \tilde{\mathcal{G}}\frac{\partial{\tilde{\mathcal{G}}}^{-1}}{\partial k_{\mu}}
  \tilde{\mathcal{G}}\frac{\partial{\tilde{\mathcal{G}}}^{-1}}{\partial k_{\nu}}
  \tilde{\mathcal{G}}\frac{\partial{\tilde{\mathcal{G}}}^{-1}}{\partial k_{\rho}}
  \right]\frac{\text{d}^3k}{(2\pi)^2},
  \label{chern_number_1}
\end{eqnarray}
where $\tilde{\mathcal{G}}(k) = \int e^{\mathtt{i}k\cdot{r}}\mathcal{G}(r)\text{d}^3r$ is
the Matsubara single-particle Green's function, with $k=(\mathtt{i}\omega,k_x,k_y)$,
$r=(-\mathtt{i}\bar{t},x,y)$, $\bar{t}$ the proper time variable,
$\mu,\nu,\rho\in\{0,1,2\}$ the coordinate indices, and $\epsilon_{\mu\nu\rho}$ being the
Levi-Civita symbol. Note that the Einstein summation convention has been applied. The
single-particle Green's function is defined as $\mathcal{G}_{ab}(r-r') \equiv
-\langle{\hat{\mathcal{T}}\big[\hat{\Psi}_{a}(r)
\hat{\Psi}^{\dagger}_{b}(r')\big]}\rangle$, with $\hat{\mathcal{T}}$ the time-ordering
operator. The Fourier transform of the Green's function $\tilde{\mathcal{G}}(k)
=\tilde{\mathcal{G}}_{\mathbf{k}} (\mathtt{i}\omega)$ can be solved by
$\tilde{\mathcal{G}}_{\mathbf{k}}(\mathtt{i}\omega) =
\big[\mathtt{i}\omega-\tilde{H}_{\mathbf{k}}
-\tilde{\Sigma}_{\mathbf{k}}(\mathtt{i}\omega)\big]^{-1}$, where the self-energy
$\tilde{\Sigma}_{\mathbf{k}}(\mathtt{i}\omega)$ includes the effect of interactions.

Hartree-Fock approximation can be used to find the band structure and solve the
single-particle Green's function of interacting systems. The field creation and
annihilation operators can be transformed as $\hat{\Psi}_{a}(\mathbf{r}) =
\sum_{n\mathbf{k}} \psi_{a,n\mathbf{k}}(\mathbf{r}) \hat{d}_{n\mathbf{k}}$ and
$\hat{\Psi}^{\dagger}_{a}(\mathbf{r}) = \sum_{n\mathbf{k}}
\psi^*_{a,n\mathbf{k}}(\mathbf{r}) \hat{d}^{\dagger}_{n\mathbf{k}}$, where
$\psi_{a,n\mathbf{k}}(\mathbf{r})$ is the hole wavefunction and
$\hat{d}^{\dagger}_{n\mathbf{k}}$/$\hat{d}_{n\mathbf{k}}$ is hole creation/annihilation
operator with band index $n$ and momentum $\mathbf{k}$. The Hartree-Fock variational
ground state for the hole-filled insulator is assumed to be $|{\text{HF}}\rangle
=\prod_{n\mathbf{k}}\hat{d}^{\dagger}_{n\mathbf{k}}|{0}\rangle$, with the vacuum state
$|{0}\rangle$ being fully-occupied valence bands (empty hole bands). A plane-wave basis
function method $\psi_{a,n\mathbf{k}}(\mathbf{r})
=\sum_{\mathbf{G}}u_{(a,\mathbf{G}),n\mathbf{k}} \phi_{\mathbf{G},\mathbf{k}}(\mathbf{r})$
can be used to find the single-particle wavefunction under Hartree-Fock approximation,
where the wavefunction coefficient $u_{(a,\mathbf{G}),n\mathbf{k}}$ is solved from a
Hartree-Fock equation $\tilde{F}_{\mathbf{k}}u_{n\mathbf{k}}
=\varepsilon_{n\mathbf{k}}u_{n\mathbf{k}}$, with $\tilde{F}_{\mathbf{k}}$ being the Fock
matrix (see Appendix \ref{sec:plane_wave} and \ref{sec:HF_band}). To study the effect of
Coulomb interaction on band topology, the band structure can be calculated by Hartree-Fock
approximation and the Chern number can be found by solving Eq.~(\ref{chern_number_1}) with
the self-energy $\tilde{\Sigma}_{\mathbf{k}}(\mathtt{i}\omega) =\tilde{F}_{\mathbf{k}}
-\tilde{H}_{\mathbf{k}}$. Under this scheme, the Fock matrix can be viewed as an effective
single-particle Hamiltonian.

\subsection{N\'eel order and gap opening\label{sec:gap_opening}}

\begin{figure}
  \includegraphics[width=0.95\linewidth]{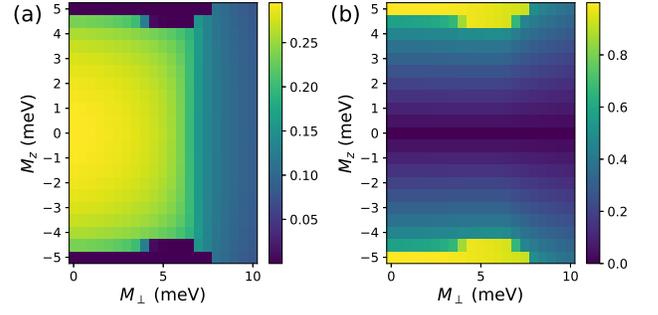}
  \caption{Color plots of (a) the $120^{\circ}$-N\'eel order parameter $\eta'$ and (b) the
  degree of valley polarization $\delta=|n_{+}-n_{-}|/(n_{+}+n_{-})$ as the functions of
  in-plane ferromagnetic magnetization $M_{\perp}$ and out-of-plane field-induced
  magnetization $M_z$ simulated by Hartree-Fock approximation of the six-band model with
  contact potential.}
  \label{fig:polarization}
\end{figure}

Since the interlayer-tunneling coupling is much smaller than the valence-band energy
offset and the holes largely reside at the MoTe${}_2$ layer, only the moir\'e band
structure of the MoTe${}_2$ layer is considered in this section. The band structure is
obtained by solving the eigenvalue problem $\tilde{F}_{\mathbf{k}}u_{n\mathbf{k}}
=\varepsilon_{n\mathbf{k}}u_{n\mathbf{k}}$. The Fock matrix is given by
$\tilde{F}_{\mathbf{k}} =\tilde{H}_{\mathbf{k}} +\tilde{K}_{\mathbf{k}}
-\tilde{J}_{\mathbf{k}}$, where $\tilde{H}_{\mathbf{k}}$ is the Bloch Hamiltonian,
$\tilde{K}_{\mathbf{k}}$ is the Coulomb-integral matrix and $\tilde{J}_{\mathbf{k}}$ is
the exchange-integral matrix (see Appendix~\ref{sec:HF_band}). The Bloch Hamiltonian in
the plane-wave basis reads $\langle \tau_1,\mathbf{G}_1| \tilde{H}_{\mathbf{k}}|
\tau_2,\mathbf{G}_2 \rangle = \int\phi^*_{\mathbf{G}_1,\mathbf{k}}(\mathbf{r})
H_{\tau_1\tau_2}(\mathbf{r}) \phi_{\mathbf{G}_2,\mathbf{k}}(\mathbf{r})\text{d}^2r$ with
$\phi_{\mathbf{G},\mathbf{k}}(\mathbf{r})= e^{\mathtt{i}\left(\mathbf{k}-\mathbf{G}\right)
\cdot\mathbf{r}}/\sqrt{S}$, where ${S}=({\sqrt{3}}/{2})Na^2_{\text{M}}$ is the area of the
moir\'e lattice with $N$ the number of moir\'e unit cell. Considering that the continuum
model for the MoTe${}_2$ monolayer is expanded by six plane-wave basis functions,
$|\tau,\mathbf{0}\rangle$, $|\tau,-\tau\bsym{g}_1\rangle$, $|\tau,\tau\bsym{g}_3\rangle$
with $\tau=\pm$, which can also be written as $|\tau,\tau\bsym{\kappa}_{J}
-\tau\bsym{\kappa}_1\rangle$ with $J=1,3,5$ and $\tau=\pm$, a six-band model can be
derived. The Hamiltonian matrix of the six-band model is written as
\begin{eqnarray}
  \tilde{H}_{\mathbf{k}}
  =
  \begin{pmatrix}
  \tilde{h}_{\mathbf{k}}-M_{z}\mathbb{I} & \tilde{\gamma}_{\mathbf{k}}\\
  \tilde{\gamma}^*_{\mathbf{k}} & \tilde{h}^*_{-\mathbf{k}}+M_{z}\mathbb{I}
  \end{pmatrix},
\end{eqnarray}
where $\mathbb{I}$ is a three-by-three identity matrix, $\tilde{h}_{\mathbf{k}}$ is the
valley-subspace Hamiltonian matrix, and $\tilde{\gamma}_{\mathbf{k}}$ is the in-plane
exchange-field matrix. The valley-subspace Hamiltonian matrix is given by
\begin{eqnarray}
  \tilde{h}_{\mathbf{k}}
  =
  \begin{pmatrix}
  \frac{|\mathbf{k}-\bsym{\kappa}_1|^2}{2m_1} & -\mathtt{i}V & \mathtt{i}V\\
    \mathtt{i}V & \frac{|\mathbf{k}-\bsym{\kappa}_3|^2}{2m_1} & -\mathtt{i}V\\
    -\mathtt{i}V & \mathtt{i}V & \frac{|\mathbf{k}-\bsym{\kappa}_5|^2}{2m_1}
  \end{pmatrix}.
\end{eqnarray}
The in-plane exchange-field matrix is written as
\begin{eqnarray}
  \tilde{\gamma}_{\mathbf{k}}
  =
  M_{\perp}
  \begin{pmatrix}
  0 & 1 & 1\\
  1 & 0 & 1\\
  1 & 1 & 0
  \end{pmatrix}
  +
  M'_{\perp}
  \begin{pmatrix}
  0 & 1 & e^{\mathtt{i}2\pi/3}\\
  1 & 0 & e^{\mathtt{i}4\pi/3}\\
  e^{\mathtt{i}2\pi/3} & e^{\mathtt{i}4\pi/3} & 0
  \end{pmatrix},\n
\end{eqnarray}
where $M'_{\perp}=0$ is assigned since the antiferromagnetic magnetization should be
contributed from the Hartree-Fock exchange interaction. The Coulomb potential in
Eq.~(\ref{many_particle}) is approximated by a contact potential $W_{ab}(\mathbf{r})\simeq
(S/N)U\delta(\mathbf{r})$, with $U$ the contact-potential energy. The Coulomb-integral
matrix is given by $\langle\tau,\tau\bsym{\kappa}_I-\tau\bsym{\kappa}_1
|\tilde{K}_{\mathbf{k}} |\tau,\tau\bsym{\kappa}_J-\tau\bsym{\kappa}_1\rangle =
U\sum_{\tau'}\langle\bsym{\kappa}_I|\rho_{\tau'\tau'} |\bsym{\kappa}_J\rangle$ and the
exchange-integral matrix is given by
$\langle\tau,\tau\bsym{\kappa}_I-\tau\bsym{\kappa}_1|\tilde{J}_{\mathbf{k}}
|\tau',\tau'\bsym{\kappa}_J-\tau'\bsym{\kappa}_1\rangle =
U\langle\bsym{\kappa}_I|\rho_{\tau\tau'}|\bsym{\kappa}_J\rangle$, where
\begin{eqnarray}
  \langle\bsym{\kappa}_I|\rho_{\tau\tau'}|\bsym{\kappa}_J\rangle
  &\equiv&
  \frac{1}{N}
  \sum_{\mathbf{k},\mathbf{G},n}\tilde{n}_{n\mathbf{k}}
  {u}_{(\tau,\mathbf{G}+\tau\bsym{\kappa}_I-\tau\bsym{\kappa}_1),n\mathbf{k}}\n
  &&\times{u}^*_{(\tau',\mathbf{G}+\tau'\bsym{\kappa}_J-\tau'\bsym{\kappa}_1),n\mathbf{k}},
\end{eqnarray}
is the single-particle density matrix. The Fock matrix can be written as
\begin{eqnarray}
  \tilde{F}_{\mathbf{k}}
  =
  \begin{pmatrix}
  \tilde{h}_{\mathbf{k}}+U\rho_{--}-M_{z}\mathbb{I}
  & \tilde{\gamma}_{\mathbf{k}}-U\rho_{+-}\\
  \tilde{\gamma}^*_{\mathbf{k}}-U\rho_{-+}
  & \tilde{h}^*_{-\mathbf{k}}+U\rho_{++}+M_{z}\mathbb{I}
  \end{pmatrix},
\end{eqnarray}
with
\begin{eqnarray}
  \rho_{\tau\tau}
  =
  \begin{pmatrix}
  n_{\tau} & \varrho_{\tau} & \varrho^*_{\tau}\\
  \varrho^*_{\tau} & n_{\tau} & \varrho_{\tau}\\
  \varrho_{\tau} & \varrho^*_{\tau} & n_{\tau}
  \end{pmatrix},\hskip1ex
  \rho_{+-}
  =
  \begin{pmatrix}
  \xi_{1} & \eta_{5} & \eta_{3}\\
  \eta_{5} & \xi_{3} & \eta_{1}\\
  \eta_{3} & \eta_{1} & \xi_{5}
  \end{pmatrix},
\end{eqnarray}
and $\rho_{-+}=\rho_{+-}^*$. Matrix elements of the single-particle density matrix are
given by $n_{\tau}=\langle\bsym{\kappa}_I|\rho_{\tau\tau}|\bsym{\kappa}_I\rangle$,
$\xi_{I}=\langle\bsym{\kappa}_I|\rho_{+-}|\bsym{\kappa}_I\rangle$ with $I=1,3,5$, and
$\varrho_{\tau}=\langle\bsym{\kappa}_1|\rho_{\tau\tau}|\bsym{\kappa}_3\rangle$,
$\eta_{1}=\langle\bsym{\kappa}_5|\rho_{+-}|\bsym{\kappa}_3\rangle$,
$\eta_{3}=\langle\bsym{\kappa}_1|\rho_{+-}|\bsym{\kappa}_5\rangle$,
$\eta_{5}=\langle\bsym{\kappa}_3|\rho_{+-}|\bsym{\kappa}_1\rangle$. If a
$120^{\circ}$-N\'eel order emerges, the exchange interaction $-U\rho_{-+}$ can be treated
as an antiferromagnetic exchange field, and the density matrix should be described by the
forms $\eta_{1}=\eta+e^{\mathtt{i}4\pi/3}\eta'$,
$\eta_{3}=\eta+e^{\mathtt{i}2\pi/3}\eta'$, $\eta_{5}=\eta+\eta'$, with $\eta$ and $\eta'$
being order parameters to quantify the ferromagnetic order and $120^{\circ}$-N\'eel order
respectively. The Hartree-Fock equation can be solved iteratively. By using the form of
the density matrix as the initial guess for the iteration of the Hartree-Fock calculation,
converged solutions of the density matrix and band structure can be obtained. A gap
opening in the moir\'e band structure can be found after the iteration. It implies that
Coulomb interaction induces a $120^{\circ}$-N\'eel order by breaking the lattice
translational symmetry and contributes to an antiferromagnetic exchange field.

A valley polarization of holes in the Hartree-Fock ground state can be induced by applying
an out-of-plane external magnetic field to the system. A degree of valley polarization for
hole doping is defined as $\delta=|n_{+}-n_{-}|/(n_{+}+n_{-})$. In
Fig.~\ref{fig:polarization}, color plots of the $120^{\circ}$-N\'eel order parameter
$\eta'$ and the degree of valley polarization $\delta$ as the functions of in-plane
ferromagnetic magnetization $M_{\perp}$ and out-of-plane magnetization $M_z$ are shown. It
is found that the $120^{\circ}$-N\'eel order and the field-induced valley polarization
compete with each other. For a wide range of magnetizations, the N\'eel order and the
valley polarization also coexist. The coexistence implies that a canted spin texture could
exist in the moir\'e superlattice. Additionally, as can be seen in
Fig.~\ref{fig:polarization} (a), the $120^{\circ}$-N\'eel order is suppressed by the
ferromagnetic magnetization but is not vanishing. It implies the coexistence of the
in-plane $120^{\circ}$-N\'eel order and the in-plane ferromagnetic order. As will be shown
later in Sec.~\ref{sec:Chern_number}, the second coexistence is crucial to the formation
of the Chern band.

\begin{figure}
  \includegraphics[width=0.9\linewidth]{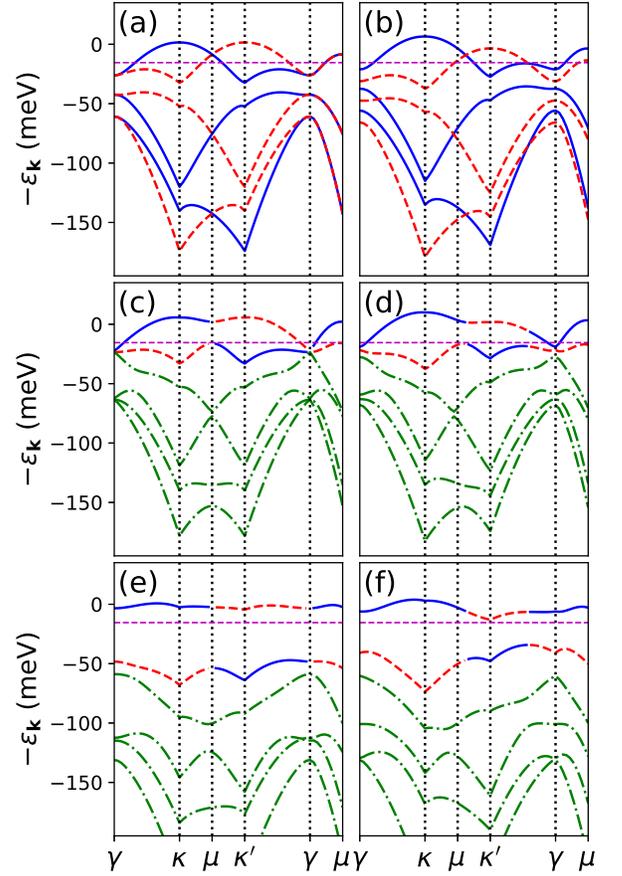}
  \caption{Moir\'e band structures contributed from the MoTe${}_2$ layer simulated by
  Hartree-Fock approximation of the six-band model with contact potential. The parameters
  by the unit of meV are (a) $M_{\perp}=0$, $M_z=0$, $U=0$; (b) $M_{\perp}=0$, $M_z=5$,
  $U=0$; (c) $M_{\perp}=10$, $M_z=0$, $U=0$; (d) $M_{\perp}=10$, $M_z=5$, $U=0$; (e)
  $M_{\perp}=10$, $M_z=0$, $U=70$; (f) $M_{\perp}=10$, $M_z=5$, $U=70$. The horizontal
  dashed lines indicate the Fermi level at $\nu$=1 filling.}
  \label{fig:HF_band}
\end{figure}

Moir\'e band structures simulated by the six-band model in the absence of Coulomb
interaction with and without the in-plane ferromagnetic exchange field are shown in
Fig.~\ref{fig:HF_band} (a), (b), (c), and (d). It is found that, in the absence of the
in-plane ferromagnetic and antiferromagnetic exchange fields, the highest two moir\'e
bands with opposite valley numbers intersect with each other. Additionally, as shown in
Fig.~\ref{fig:HF_band} (b), the crossing over between two moir\'e bands, which can be seen
as a band inversion, survives with a small out-of-plane magnetic field. The horizontal
dashed line indicates the Fermi level at $\nu$=1 filling. Since the Fermi level crosses
the moir\'e bands, holes occupying these bands would form a Fermi sea and show a metallic
transport property. In Fig.~\ref{fig:HF_band} (c) and (d), it is found that the in-plane
ferromagnetic exchange field opens a gap at the band inversion along the line $\gamma-\mu$
in the MBZ. However, the gap is not entirely opened at the high symmetry point $\gamma$
and the Fermi level is still crossing the moir\'e bands. In Fig.~\ref{fig:HF_band} (e) and
(f), moir\'e band structures solved from Hartree-Fock approximation of the six-band model
with contact potential are shown. A gap is opened along the crossing line between the two
highest bands. The gap is opened by the antiferromagnetic exchange field induced by the
$120^{\circ}$-N\'eel order. The hole-occupied band, the highest band in
Fig.~\ref{fig:HF_band} (f), has a valley-polarized population of holes majorly with
$\tau=+$. Because of the gap, the $120^{\circ}$-N\'eel order and the field-induced valley
polarization can coexist. Since the Fermi level resides between the highest two moir\'e
bands, the holes only occupy the highest moir\'e band and show an insulating transport
property. It indicates that the $120^{\circ}$-N\'eel order ensures the insulating gap.

\subsection{Chern number\label{sec:Chern_number}}

\begin{figure}
  \includegraphics[width=0.95\linewidth]{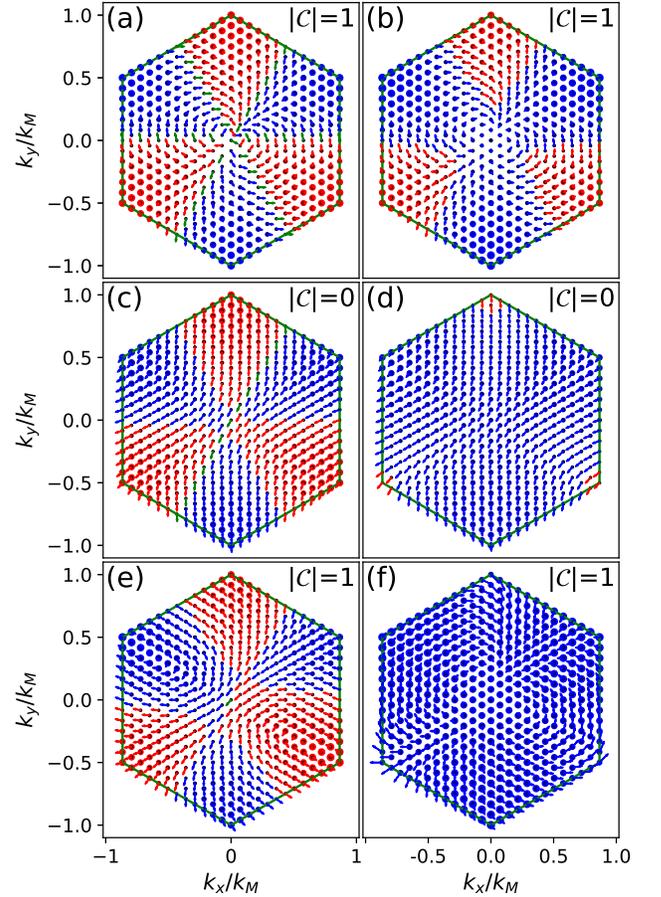}
  \caption{Fock pseudospin textures in the MBZ contributed from the MoTe${}_2$ layer
  simulated by Hartree-Fock approximation of the six-band model with contact potential.
  The parameters by the unit of meV are (a) $M_{\perp}=5$, $M_z=0$, $U=0$; (b)
  $M_{\perp}=5$, $M_z=3$, $U=0$; (c) $M_{\perp}=5$, $M_z=0$, $U=70$; (d) $M_{\perp}=5$,
  $M_z=3$, $U=70$; (e) $M_{\perp}=10$, $M_z=0$, $U=70$; (f) $M_{\perp}=5$, $M_z=5$,
  $U=70$.}
  \label{fig:moire_texture}
\end{figure}

In this section, the Chern numbers of moir\'e bands simulated by Hartree-Fock
approximation of the six-band model with contact potential are studied. To demonstrate the
emergence of the topological order, the Fock matrix is reduced to a two-by-two matrix by a
projection transformation as
\begin{eqnarray}
  \tilde{\mathcal{F}}_{\mathbf{k}}
  =
  \mathcal{P}^{\dagger}_{\mathbf{k}}\tilde{F}_{\mathbf{k}}\mathcal{P}_{\mathbf{k}}
  =
  \begin{pmatrix}
  \varepsilon_{+,\mathbf{k}}-M_{z} & f_{+-,\mathbf{k}}\\
  f_{-+,\mathbf{k}} & \varepsilon_{-,\mathbf{k}}+M_{z}
  \end{pmatrix},
\end{eqnarray}
where $\mathcal{P}_{\mathbf{k}} = \begin{pmatrix} \mathcal{P}_{+,\mathbf{k}} & 0\\ 0 &
\mathcal{P}_{-,\mathbf{k}} \end{pmatrix}$ is the projection matrix with
$\mathcal{P}_{\pm,\mathbf{k}}$ following the unitary condition
$\mathcal{P}^{\dagger}_{\pm,\mathbf{k}}\mathcal{P}_{\pm,\mathbf{k}}=1$, and
$\varepsilon_{\pm,\mathbf{k}}$ is solved from the eigenvalue equation
$\big(\tilde{h}_{\mathbf{k}}+U\rho_{\mp\mp}\big)\mathcal{P}_{\pm,\mathbf{k}}
=\varepsilon_{\pm,\mathbf{k}}\mathcal{P}_{\pm,\mathbf{k}}$ as the lowest eigenvalue. The
off-diagonal matrix elements are given by $f_{-+,\mathbf{k}} =
\mathcal{P}^{\dagger}_{-,\mathbf{k}} \big(\tilde{\gamma}^*_{\mathbf{k}}-U\rho_{-+}\big)
\mathcal{P}_{+,\mathbf{k}}$, $f_{+-,\mathbf{k}} = \mathcal{P}^{\dagger}_{-,\mathbf{k}}
\big(\tilde{\gamma}_{\mathbf{k}}-U\rho_{+-}\big) \mathcal{P}_{+,\mathbf{k}}$. The reduced
Fock matrix can be reformulated by the parametrized Fock pseudospin
$\tilde{\mathcal{F}}_{\mathbf{k}}=\tilde{\mathcal{F}}_{\mathbf{k},0}\mathbb{I}
+\tilde{\mathcal{F}}_{\mathbf{k},x}\sigma_{x}
+\tilde{\mathcal{F}}_{\mathbf{k},y}\sigma_{y}
+\tilde{\mathcal{F}}_{\mathbf{k},z}\sigma_{z}$. Based on Eq.~(\ref{chern_number_1}) and
the argument in Sec.~\ref{sec:interacting}, the Chern number can be calculated
by\cite{bernevig2013topological}
\begin{eqnarray}
  \mathcal{C}
  =
  \frac{1}{4\pi}\int_{\text{BZ}}\frac{\tilde{\bsym{\mathcal{F}}}_{\mathbf{k}}
  \cdot\partial_{k_{x}}\tilde{\bsym{\mathcal{F}}}_{\mathbf{k}}
  \times\partial_{k_{y}}\tilde{\bsym{\mathcal{F}}}_{\mathbf{k}}}
  {|\tilde{\bsym{\mathcal{F}}}_{\mathbf{k}}|^3}\text{d}^2k.
  \label{chern_number_2}
\end{eqnarray}

Illustrations of Fock pseudospin textures are shown in Fig.~\ref{fig:moire_texture} with
different sets of parameters. The blue dots indicate the $k$ points that
$\tilde{\mathcal{F}}_{\mathbf{k},z}<0$ and the red dots indicate the points that
$\tilde{\mathcal{F}}_{\mathbf{k},z}>0$. The dot size indicates the relative value of
$|\tilde{\mathcal{F}}_{\mathbf{k},z}|$, and the arrows point to the direction of
$(\tilde{\mathcal{F}}_{\mathbf{k},x}/| \bsym{\tilde{\mathcal{F}}}_{\mathbf{k}}|,\;
\tilde{\mathcal{F}}_{\mathbf{k},y}/| \bsym{\tilde{\mathcal{F}}}_{\mathbf{k}}|)$. As can be
seen, the Fock pseudospin in Fig.~\ref{fig:moire_texture} (c) and (d) show topological
trivial textures, and the Fock pseudospin in Fig.~\ref{fig:moire_texture} (a), (b), (e),
(f) show topological nontrivial skyrmion textures. The textures in
Fig.~\ref{fig:moire_texture} (c) and (d) contribute no Chern number to the moir\'e band
structures. The textures in Fig.~\ref{fig:moire_texture} (a), (b), (e), and (f) contribute
a unit Chern number to each hole-occupied band via the winding number of the skyrmion
texture in the MBZ. It is found that, as can be seen in Fig.~\ref{fig:moire_texture} (a)
and (b), topological nontrivial textures can be generated by the in-plane ferromagnetic
exchange field without the Hartree-Fock exchange interaction. The inclusion of the
exchange interaction actually could make the Fock pseudospin textures trivial, as shown in
Fig.~\ref{fig:moire_texture} (c) and (d). With a higher in-plane ferromagnetic
magnetization or a higher out-of-plane field-induced magnetization, as shown in
Fig.~\ref{fig:moire_texture} (e) and (f), the Fock pseudospin textures again become
topological nontrivial. It implies that, firstly, the topological order can be induced
solely by the ferromagnetic exchange field. Secondly, the $120^{\circ}$-N\'eel order
competes with the topological order rather than assists it. Thirdly, since the
field-induced valley polarization of holes competes with the $120^{\circ}$-N\'eel order,
the N\'eel order is reduced under an external magnetic field, and the reduction
facilitates the formation of topological nontrivial textures. However, it is needed to
note that, as discussed in Sec.~\ref{sec:gap_opening}, the $120^{\circ}$-N\'eel order
contributes to the insulating gap between the two highest bands at zero magnetic field
such that the hole-occupied state can be an insulator at $\nu=1$ hole filling. Therefore,
the $120^{\circ}$-N\'eel order is indispensable for the generation of a QAH state, even if
it also competes with the formation of the Chern band.

In short summary, the in-plane ferromagnetic order generates the Chern band, and the
in-plane $120^{\circ}$-N\'eel order induces the insulating gap in the moir\'e band
structure. Since the topological order emerges as the ferromagnetic order is formed, and
the insulating gap has been opened before and after the formation of the ferromagnetic
order, there is no charge gap closure at the topological phase transition.

\section{Exciton condensation and ferromagnetism\label{sec:trs_breaking}}

There are two unsolved problems in the current argument. Firstly, the in-plane
ferromagnetic exchange field in the discussion is artificially introduced to the model.
Secondly, the bandwidth of the hole-occupied band in the first MBZ is about
$E_W=k^2_{\text{M}}/({2m_1})\simeq 43$ meV, and the contact-potential energy $U=70$ meV
can be seen as the on-site Coulomb repulsion within a moir\'e unit cell. Since $U>E_W$,
the equilibrium state should be a Mott-insulator state, and thus the band-structure
picture to describe the electronic structure is artificial. To solve these problems, we
suggest that an interlayer-exciton condensate is formed at $\nu=1$ filling under the
out-of-plane electric field. An in-plane ferromagnetic order is generated by the
equilibrium exciton condensate via a mechanism called exciton ferromagnetism. At a certain
electric field, a correlated insulating state composed of the exciton condensate and the
hole-occupied band becomes a new thermodynamically stable phase. Therefore, the
band-structure picture can still be available, and a topological phase transition could
occur as the ferromagnetic order emerges. In this section, descriptions of exciton
condensation and ferromagnetism are provided. In Sec.~\ref{sec:gap_equation}, the
Hamiltonian for studying the interlayer-exciton condensate and the gap equation for the
exciton order parameter are introduced. In Sec.~\ref{sec:survival}, the survival of the
Chern band in the presence of the exciton condensate is discussed. In
Sec.~\ref{sec:exciton_ferromagnetism}, the theory of exciton ferromagnetism is introduced.
In Sec.~\ref{sec:exciton_mott}, we argue that the observed insulator-to-metal transition
at a higher electric field can be attributed to an exciton Mott transition. Some
derivations and formulations of the gap equation, exciton binding energy, and exciton
instability for exciton condensation are given in Appendix~\ref{sec:exciton_condensation}.

\subsection{Interlayer-exciton condensate\label{sec:gap_equation}}

Exciton condensation is Bose-Einstein condensation (BEC) of
excitons\cite{jerome1967excitonic, zittartz1967theory, halperin1968excitonic,
keldysh1968collective, comte1982exciton, nozieres1982exciton, fernandez1997spin,
chu1996theory, wu2015theory}. Interlayer-exciton condensates have been observed in layered
materials\cite{li2017excitonic, wang2019evidence, ma2021strongly, gu2022dipolar,
chen2022excitonic, zhang2022correlated}. An interlayer-exciton condensate (with the
intralayer Coulomb repulsion being omitted, which will be discussed later) can be studied
by the electron-hole-system (EHS) Hamiltonian
\begin{eqnarray}
  \hat{\mathcal{H}}_{\text{EHS}}
  &=&
  \sum_{\mathbf{k}}\varepsilon^{\text{e}}_{\mathbf{k}}
  \hat{c}^{\dagger}_{\mathbf{k}}\hat{c}_{\mathbf{k}}
  +
  \sum_{\tau,\mathbf{k}}\varepsilon^{\text{h}}_{\tau,\mathbf{k}}
  \hat{d}^{\dagger}_{\tau,\mathbf{k}}\hat{d}_{\tau,\mathbf{k}}\n
  &&
  -\sum_{\tau,\mathbf{q}\mathbf{k}\mathbf{k}'}
  \frac{W^{\text{eh}}_{\mathbf{q}}}{S}
  \hat{c}^{\dagger}_{\mathbf{k}-\mathbf{q}}
  \hat{d}^{\dagger}_{\tau,\mathbf{k}'+\mathbf{q}}
  \hat{d}_{\tau,\mathbf{k}'}
  \hat{c}_{\mathbf{k}},
  \label{eh_Hamiltonian}
\end{eqnarray}
where $\hat{c}^{\dagger}_{\mathbf{k}}$/$\hat{c}_{\mathbf{k}}$ is the electron
creation/annihilation operator on the unfilled valence (hole-occupied) band in the
MoTe${}_2$ layer, (the highest band of Fig.~\ref{fig:HF_band} (e))
$\hat{d}^{\dagger}_{\tau,\mathbf{k}}$/$\hat{d}_{\tau,\mathbf{k}}$ is the hole
creation/annihilation operator on the two-component valence band in the WSe${}_2$ layer
with $\tau=\pm$, (the two highest bands in Fig.~\ref{fig:moire_band} (b)),
$\varepsilon^{\text{e}}_{\mathbf{k}}$ and $\varepsilon^{\text{h}}_{\tau,\mathbf{k}}$ are
electron and hole band energies, $W^{\text{eh}}_{\mathbf{q}}$ is the interlayer
electron-hole interaction, and $S$ is the area of the system. The band energies are given
by
\begin{eqnarray}
  \varepsilon^{\text{e}}_{\mathbf{k}}
  \simeq
  0,\hskip2ex
  \varepsilon^{\text{h}}_{\tau,\mathbf{k}}
  \simeq
  D-\xi_{z}\mathcal{F}_{z}+\frac{k^2}{2m_2},
\end{eqnarray}
where $\xi_{z}=2.6$ e$\cdot$\AA\; is the interlayer dipole moment\cite{li2021continuous},
$\mathcal{F}_{z}$ is the out-of-plane electric field, and the hole-band dispersion is
subject to an upper limit cutoff momentum $k_{\text{cutoff}}=k_{\text{M}}$. The electron
band is assumed to be flatten as shown in Fig~\ref{fig:HF_band} (e), (f). The interlayer
electron-hole interaction is given by the modified Rytova-Keldysh potential
${W}^{\text{eh}}_{\mathbf{q}} = {2\pi}/[{\epsilon({q})q}]$, with\cite{rytova, keldysh,
van2018interlayer}
\begin{eqnarray}
  \epsilon({q})
  &=&
  [\epsilon_{\text{r}}+(\rho_1+\rho_2)q/2]\cosh{dq}\n
  &&+[1+(\epsilon_{\text{r}}+\rho_1q)(\epsilon_{\text{r}}+\rho_2q)/(4\epsilon_{\text{r}})]
  \sinh{dq},
\end{eqnarray}
where $\epsilon_{\text{r}}=4.0$ is the dielectric constant for the surrounding hexagonal
boron nitride layers, $\rho_{l}$ is the screening length on the $l$-th layer, with
$\rho_{1}=73.61$ \AA\; for the MoTe${}_2$ layer and $\rho_{2}=47.57$ \AA\; for the
WSe${}_2$ layer\cite{kylanpaa2015binding}, and $d=7.0$ \AA\; is the interlayer distance.
The exciton binding energy ($E_{\text{X}}$) can be obtained by solving
\begin{eqnarray}
  \left(\frac{k^2}{2m_2}+E_{\text{X}}\right)\tilde{\Psi}_{{\tau},\mathbf{k}}
  =
  \sum\nolimits_{\mathbf{k}'}\frac{W^{\text{eh}}_{\mathbf{k}-\mathbf{k}'}}{S}
  \tilde{\Psi}_{{\tau},\mathbf{k}'}
\end{eqnarray}
variationally\cite{mypaper0} (also see Appendix~\ref{sec:exciton_binding_energy}). It is
found to be $E_{\text{X}}=129$ meV and the projected in-plane exciton radius is
$a_{\text{X}}=16$ \AA. An exciton condensate can be formed if the exciton binding energy
is larger than the band gap\cite{jerome1967excitonic, keldysh1968collective,
comte1982exciton} (also see Appendix~\ref{sec:excitonic_instability}), indicating
$E_{\text{X}}>\tilde{D}$ with $\tilde{D}=D-\xi_{z}\mathcal{F}_{z}$ the reorganized band
gap. While $\mathcal{F}_{z}=0.66$ V/nm and $\tilde{D}=108$ meV at the topological phase
transition\cite{exp0}, the condition is satisfied.

\begin{figure}
  \includegraphics[width=0.95\linewidth]{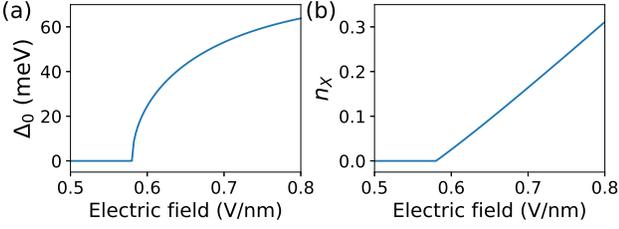}
  \caption{Gap-equation calculation. (a) Exciton order parameter
  $\Delta_{0}=\Delta_{\tau,\mathbf{k}=\mathbf{0}}$ and (b) exciton density $n_{\text{X}}$
  v.s. electric-field strength $\mathcal{F}_{z}$.}
  \label{fig:exciton}
\end{figure}

The equilibrium exciton condensate can be described by the Bardeen-Cooper-Schrieffer
(BCS)-like wavefunction
\begin{eqnarray}
  |{\Phi_{\text{BCS}}}\rangle = \prod_{\tau,\mathbf{k}}
  \big(u_{\tau,\mathbf{k}}+v_{\tau,\mathbf{k}}\hat{d}^{\dagger}_{\tau,\mathbf{k}}
  \hat{c}^{\dagger}_{-\mathbf{k}}\big)|{\Phi_{\text{HF}}}\rangle,
\end{eqnarray}
where $u_{\tau,\mathbf{k}}$ and $v_{\tau,\mathbf{k}}$ are variational coefficients subject
to the normalization condition
$u_{\tau,\mathbf{k}}^2+v_{\tau,\mathbf{k}}^2=1$\cite{jerome1967excitonic,
comte1982exciton}. The variational coefficients can be solved as $u^2_{\tau,\mathbf{k}} =
\left(1+{{\Xi}_{\tau,\mathbf{k}}} /{\mathcal{E}_{\tau,\mathbf{k}}}\right)/2$,
$v^2_{\tau,\mathbf{k}} = \left(1-{{\Xi}_{\tau,\mathbf{k}}}
/{\mathcal{E}_{\tau,\mathbf{k}}}\right)/2$, where $\mathcal{E}_{\tau,\mathbf{k}} =
\sqrt{|{\Xi}_{\tau,\mathbf{k}}|^2+|\Delta_{\tau,\mathbf{k}}|^2}$ and
${\Xi}_{\tau,\mathbf{k}} = \big(\varepsilon^{\text{e}}_{-\mathbf{k}}
+\varepsilon^{\text{h}}_{\tau,\mathbf{k}} \big)/2$. The exciton density is given by
$n_{\text{X}} =\sum_{\tau,\mathbf{k}}v^2_{{\tau},\mathbf{k}}/N$, and the exciton order
parameter $\Delta_{\tau,\mathbf{k}}$ can be solved from the gap equation (also see
Appendix~\ref{sec:variation})
\begin{eqnarray}
  \Delta_{\tau,\mathbf{k}} =
  \frac{1}{2S}\sum_{\mathbf{k}'}\frac{W^{\text{eh}}_{\mathbf{k}-\mathbf{k}'}
  \Delta_{\tau,\mathbf{k}'}}{\mathcal{E}_{\tau,\mathbf{k}'}}.
  \label{gap_eq_1}
\end{eqnarray}
The results of the gap-equation calculation are shown in Fig.~\ref{fig:exciton}. The
exciton condensate is formed at the electric-field strength of about $0.58$ V/nm, which is
lower than the observed value $0.66$ V/nm at the topological phase transition\cite{exp0}.
The additional electric-field strength is required for the equilibrium exciton condensate
to become a new stable state by replacing the Mott-insulator state.

\subsection{Survival of the Chern band\label{sec:survival}}

Since the hole-occupied moir\'e band is the Chern band in the $\text{MoTe}_2$ layer, it is
essential to be aware of the possibility that the topology of the hole-occupied moir\'e
band could be altered in the presence of the exciton condensate. To clarify that, we
consider the quasiparticle Green's function for the exciton
condensate\cite{jerome1967excitonic, zittartz1967theory, chu1996theory}
\begin{eqnarray}
  {\mathcal{G}}_{\tau,\mathbf{k}}(\bar{t})
  &\equiv&
  -
  \begin{pmatrix}
  \langle{\hat{\mathcal{T}}\hat{c}_{-\mathbf{k}}(0)
  \hat{c}^{\dagger}_{-\mathbf{k}}(\bar{t})}\rangle
  & \langle{\hat{\mathcal{T}}\hat{c}^{\dagger}_{-\mathbf{k}}(\bar{t})
  \hat{d}^{\dagger}_{\tau,\mathbf{k}}(0)}\rangle\\
  \langle{\hat{\mathcal{T}}\hat{d}_{\tau,\mathbf{k}}(\bar{t})
  \hat{c}_{-\mathbf{k}}(0)}\rangle
  & \langle{\hat{\mathcal{T}}\hat{d}_{\tau,\mathbf{k}}(\bar{t})
  \hat{d}^{\dagger}_{\tau,\mathbf{k}}(0)}\rangle
  \end{pmatrix}.\n
  \label{quasiparticle_green_function}
\end{eqnarray}
By a Fourier transform $\mathcal{G}_{\tau,\mathbf{k}}(\mathtt{i}\omega_\nu)
=\frac{1}{\beta}\int^{\beta}_{0} e^{\mathtt{i}\omega_\nu\bar{t}}
{\mathcal{G}}_{\tau,\mathbf{k}}(\bar{t})\text{d}\bar{t}$, with $\beta$ the inverse
temperature and $\omega_\nu=(2\nu+1)\pi/\beta$, the quasiparticle Green's function can be
solved as
\begin{eqnarray}
  \tilde{\mathcal{G}}^{-1}_{\tau,\mathbf{k}}(\mathtt{i}\omega)
  &=&
  \begin{pmatrix}
  \mathtt{i}\omega+\varepsilon^{\text{e}}_{-\mathbf{k}} & -\Delta_{\tau,\mathbf{k}}\\
  -\Delta_{\tau,\mathbf{k}} & \mathtt{i}\omega
  -\varepsilon^{\text{h}}_{\tau,\mathbf{k}}
  \end{pmatrix}.
  \label{quasiparticle_qf2}
\end{eqnarray}
The electron creation and annihilation operators can be replaced by the hole creation and
annihilation operators by $\hat{c}^{\dagger}_{-\mathbf{k}} =\hat{d}_{\mathbf{k}}$,
$\hat{c}_{-\mathbf{k}} =\hat{d}^{\dagger}_{\mathbf{k}}$. With including the valence bands
in the $\text{MoTe}_2$ layer, the quasiparticle Green's function can be generalized by the
formulation ${\mathcal{G}}_{nm,\mathbf{k}}(\bar{t}) \equiv -\langle{\hat{\mathcal{T}}
\hat{d}_{n\mathbf{k}}(\bar{t}) \hat{d}^{\dagger}_{m\mathbf{k}}(0)}\rangle$, with $n$, $m$
indexing different valence bands. This hole Green's function satisfies a Ward-Takahashi
identity\cite{chu1996theory}, and thus the Chern number of the hole bands can also be
calculated by Eq.~(\ref{chern_number_1}). Therefore, based on the Green's function given
in Eq.~(\ref{quasiparticle_qf2}), the effect of forming an exciton condensate on the band
structure can be realized as the hybridization between the unfilled valence band
(hole-occupied band) in the $\text{MoTe}_2$ layer and the valence bands in the
$\text{WSe}_2$ layer. The topology of the Chern band will be altered only if
$2{\Xi}_{\tau,\mathbf{k}}\leq\Delta_{\tau,\mathbf{k}}$, in which a band inversion could
occur. Since $2{\Xi}_{\tau,\mathbf{k}}>\Delta_{\tau,\mathbf{k}}$ can be ensured by
$\tilde{D}>\Delta_{0}$ as shown in Fig.~\ref{fig:exciton} (a), the survival of the Chern
band with the exciton condensate is ensured.

\subsection{Exciton ferromagnetism\label{sec:exciton_ferromagnetism}}

In this section, we argue that an in-plane ferromagnetic order can be induced by exciton
condensation and exciton-exciton interaction, and the in-plane ferromagnetic exchange
field in the continuum model is contributed from the ferromagnetic order. Note that the
moir\'e periodicity and the intralayer Coulomb repulsion have not been considered in the
EHS Hamiltonian in Eq.~(\ref{eh_Hamiltonian}). The moir\'e periodicity and the Coulomb
repulsion can lead to the localization of an exciton in each moir\'e unit cell. Such an
effect on the exciton condensate can be described by the excitonic Bose-Hubbard (EBH)
Hamiltonian\cite{lagoin2021key, remez2021dark, gotting2022moir},
\begin{eqnarray}
  \hat{\mathcal{H}}_{\text{EBH}}
  &=&
  -t\sum_{\tau,\langle{\mathbf{R},\mathbf{R}'}\rangle}
  \hat{x}^{\dagger}_{\tau,\mathbf{R}}\hat{x}_{\tau,\mathbf{R}'}
  +U\sum_{\mathbf{R}}\hat{x}^{\dagger}_{+,\mathbf{R}}
  \hat{x}^{\dagger}_{-,\mathbf{R}}\hat{x}_{-,\mathbf{R}}\hat{x}_{+,\mathbf{R}}\n
  &&+U'\sum_{\tau,\mathbf{R}}\hat{x}^{\dagger}_{\tau,\mathbf{R}}
  \hat{x}_{\tau,\mathbf{R}}(\hat{x}^{\dagger}_{\tau,\mathbf{R}}\hat{x}_{\tau,\mathbf{R}}-1),
\end{eqnarray}
where $\hat{x}^{\dagger}_{\tau,\mathbf{R}} =
\frac{1}{\sqrt{N}}\sum_{\mathbf{K}}{e^{-\mathtt{i}\mathbf{K}\cdot\mathbf{R}}}
\hat{X}^{\dagger}_{\tau,\mathbf{K}}$ is the exciton creation operator on the moir\'e unit
cell at $\mathbf{R}$ site, with $\hat{X}^{\dagger}_{\tau,\mathbf{K}}
=\sum_{\mathbf{k}}{\Psi}_{\tau,\mathbf{k},\mathbf{K}}
\hat{d}^{\dagger}_{\tau,\mathbf{k}+\mathbf{K}} \hat{c}^{\dagger}_{-\mathbf{k}}$ and
${\Psi}_{\tau,\mathbf{k},\mathbf{K}}$ the wavefunction for the interlayer exciton in the
moir\'e potential, $t$ is the nearest-neighbor hopping coupling, $U$ is the intervalley
on-site repulsion, and $U'$ is the intravalley on-site repulsion. The filling number of
each moir\'e unit cell is given by the exciton density $n_{\text{X}}$. Since the exciton
condensate is a BEC, the condensate can be assumed to be fragmented and distributed
equally throughout the moir\'e superlattice\cite{mueller2006fragmentation,
ueda2010fundamentals}. The exciton condensate can then be described by a rescaled EBH
Hamiltonian with $\hat{x}_{\tau,\mathbf{R}}\rightarrow
\sqrt{n_{\text{X}}}\hat{\tilde{x}}_{\tau,\mathbf{R}}$, $t\rightarrow\tilde{t}$,
$U\rightarrow\tilde{U}/n_{\text{X}}$, $U'\rightarrow\tilde{U}'/{n}_{\text{X}}$,
$\hat{\mathcal{H}}_{\text{EBH}}\rightarrow{n}_{\text{X}}
\hat{\tilde{\mathcal{H}}}_{\text{EBH}}$ and the filling of the moir\'e superlattice
becomes one exciton per unit cell. If $|\tilde{t}|\ll \tilde{U},\tilde{U}'$, the EBH
Hamiltonian with the filling number being one can be approximated by the anisotropic
Heisenberg (XXZ) Hamiltonian\cite{kuklov2003counterflow, altman2003phase,
duan2003controlling, he2012quantum}
\begin{eqnarray}
  \hat{\mathcal{H}}_{\text{XXZ}}
  =
  \sum_{\langle{\mathbf{R},\mathbf{R}'}\rangle}
  \left[J_{z}\hat{\mathcal{S}}^{z}_{\mathbf{R}}\hat{\mathcal{S}}^{z}_{\mathbf{R}'}
  -J_{\perp}
  \left(\hat{\mathcal{S}}^{x}_{\mathbf{R}}\hat{\mathcal{S}}^{x}_{\mathbf{R}'}
  +\hat{\mathcal{S}}^{y}_{\mathbf{R}}\hat{\mathcal{S}}^{y}_{\mathbf{R}'}\right)\right],\n
\end{eqnarray}
with $J_{z}=4\tilde{t}^2/\tilde{U}-4\tilde{t}^2/\tilde{U}'$,
$J_{\perp}=4\tilde{t}^2/\tilde{U}$ and $\hat{\mathcal{S}}^{x,y,z}_{\mathbf{R}}$ being
pseudospin operators spanned by the basis $|{\tau}=\pm\rangle$. The model exhibits a
transverse ferromagnetic order if $J_{\perp}>J_{z}$\cite{he2012quantum,
kosterlitz1973ordering}. The ferromagnetic order induces TRS breaking and the in-plane
ferromagnetic exchange field in the continuum model. It is estimated that $U\simeq
U'\simeq e^2/(\epsilon_{\text{r}} a_{\text{M}})\simeq 70$ meV, $t\simeq 0.015U\simeq 1$
meV\cite{gotting2022moir}, and $n_{\text{X}}\simeq 0.10$ at the QAH state. We get
$\tilde{U}=Un_{\text{X}}\simeq 7.0$ meV, $J_{\perp}\simeq 0.63$ meV, and $J_{z}\simeq{0}$
meV. The Berezinskii-Kosterlitz-Thouless (BKT) temperature\cite{kosterlitz1973ordering} in
a triangular lattice is estimated to be about $T_{\text{BKT}}\simeq
(1/0.69)J_{\perp}/(2k_{\text{B}})$\cite{butera1994high}, which gives $T_{\text{BKT}}\simeq
5$ K. This scale is consistent with the observed Curie temperature for the ferromagnetic
transition\cite{exp0}. The ferromagnetic exchange field can be obtained from the
mean-field approximation of the XXZ model
\begin{eqnarray}
  \hat{\mathcal{H}}_{\text{XXZ}}
  &\simeq&
  (J_{\perp}/2)\sum_{\langle{\mathbf{R},\mathbf{R}'}\rangle}
  \left(\langle\hat{\mathcal{S}}^{+}_{\mathbf{R}}\rangle
  \langle\hat{\mathcal{S}}^{-}_{\mathbf{R}'}\rangle
  +\langle\hat{\mathcal{S}}^{-}_{\mathbf{R}}\rangle
  \langle\hat{\mathcal{S}}^{+}_{\mathbf{R}'}\rangle\right)\n
  &&-J_{\perp}\sum_{\langle{\mathbf{R},\mathbf{R}'}\rangle}
  \left(\hat{\mathcal{S}}^{+}_{\mathbf{R}}
  \langle\hat{\mathcal{S}}^{-}_{\mathbf{R}'}\rangle
  +\hat{\mathcal{S}}^{-}_{\mathbf{R}}
  \langle\hat{\mathcal{S}}^{+}_{\mathbf{R}'}\rangle\right),
\end{eqnarray}
with $\hat{\mathcal{S}}^{\pm}_{\mathbf{R}}
=\hat{\mathcal{S}}^{x}_{\mathbf{R}}\pm\mathtt{i}\hat{\mathcal{S}}^{y}_{\mathbf{R}}
=\hat{x}^{\dagger}_{\pm,\mathbf{R}}\hat{x}_{\mp,\mathbf{R}}$. The in-plane ferromagnetic
magnetization can be estimated by $M_{\perp}\simeq zJ_{\perp}$ with $z$ the coordination
number and $z=6$ for triangular lattices. The value is estimated to be $M_{\perp}\simeq
3.8$ meV. This value of the ferromagnetic magnetization is at the same scale as our
assumed values in the calculations in Sec.~\ref{sec:Chern_number}, but it is smaller than
the value required for the topological nontrivial texture in Fig.~\ref{fig:moire_texture}
(e) to show up without a valley polarization of holes. The discrepancy might be caused by
the roughness of the present estimation or the lack of considering long-range interaction
for both the Hartree-Fock calculation and exciton ferromagnetism. We will return to the
topic in future studies to improve the estimation.

\subsection{Exciton Mott transition\label{sec:exciton_mott}}

With a denser population of excitons under a higher electric field, the
correlation-induced screening effect and Pauli-blocking effect may cause the dissociation
of excitons and the formation of electron-hole plasma. The mechanism is known as exciton
Mott transition\cite{hanamura1977condensation, keldysh1986electron,
zimmermann1988nonlinear, asano2014exciton, fogler2014high, rustagi2018theoretical}. A Mott
density $\rho_{\text{Mott}}$ is defined as the critical exciton density in which the
dissociation of excitons occurs. Two different theoretical schemes have been proposed to
estimate the Mott density. One theoretical scheme suggests that the Mott density is the
exciton density in which exciton wavefunctions begin to overlap mutually. For
two-dimensional systems, the Mott density is estimated to be
$\rho_{\text{Mott}}a^2_{\text{X}}\simeq 0.3\sim 0.7$. This scheme has been supported by
using quantum Monte-Carlo calculations\cite{de2002excitonic, rios2018evidence}. The other
theoretical scheme assumes that the Mott density is reached as the
electron-hole-excitation-induced band-gap renormalization energy is larger than the
exciton binding energy. The Mott density estimated by this
scheme\cite{hanamura1977condensation, zimmermann1988nonlinear, asano2014exciton,
rustagi2018theoretical} is about $\rho_{\text{Mott}}a^2_{\text{X}}\simeq 0.02\sim 0.08$.
While both theoretical schemes have gathered supporters, recent experiments on exciton
Mott transition in the moir\'e-bilayer system suggest
$\rho_{\text{Mott}}a^2_{\text{X}}\simeq 0.01\sim 0.07$\cite{wang2021diffusivity,
siday2022ultrafast}, which strongly supports the second scheme.

For MoTe${}_2$/WSe${}_2$ moir\'e heterobilayers, if $\rho_{\text{Mott}}a^2_{\text{X}}
\simeq 0.02$ is assumed, the exciton density per moir\'e unit cell at the exciton Mott
transition is estimated to be $n_{\text{X}}=\rho_{\text{Mott}}\bar{S}\simeq
0.02\times(\sqrt{3}/2)\times a^2_{\text{M}}/a^2_{\text{X}}\simeq 0.17$, which can be
reached by the imposed electric-field strength $0.70$ V/nm according to
Fig.~\ref{fig:exciton}. (b). This estimation is consistent with the observed electric
field in which the metallic phases show up in both AA-stacked and AB-stacked
MoTe${}_2$/WSe${}_2$ heterobilayers\cite{exp0, li2021continuous}. At low temperatures, the
exciton Mott transition is a quantum phase transition between BEC-like exciton-gas
condensation and BCS-like electron-hole-liquid condensation\cite{bronold2006possibility,
kremp2008quantum}. Since the BEC-to-BCS transition is known as a continuous
crossover\cite{bronold2006possibility, kremp2008quantum} and the electron-hole-liquid
condensate can also be seen as a two-component Fermi
liquid\cite{hanamura1977condensation}, the observed continuous insulator-to-metal
transition in MoTe${}_2$/WSe${}_2$ heterobilayers at low temperatures\cite{exp0,
li2021continuous} could be explained by the exciton Mott transition.

\section{Discussions and Conclusion\label{sec:discussion}}

The consistency between the present theory and experimental observations is discussed
sequentially regarding the enumerated list in the introduction:
\begin{enumerate}
\item The continuum model gives the bandwidth $E_{W}=43$ meV for the highest moir\'e band
in the MoTe${}_2$ layer and the contact-interaction energy $U=70$ meV for the on-site
Coulomb repulsion. Since $U>E_{W}$, the equilibrium state is a Mott-insulator state in low
electric fields. An interlayer-exciton condensate is formed at $\nu=1$ hole filling and a
certain electric field. A correlated insulating state composed of the hole-occupied band
and the exciton condensate becomes a new stable phase while competing with the
Mott-insulator state, such that the band-structure picture can still be available beyond
the electric-field strength.
\item The valence-band energy offset is assumed to be $D=280$ meV, which is not far from
the observed value of $300$ meV. At the topological phase transition, the valence-band
energy offset is reduced to $\tilde{D}=108$ meV, which is still much larger than the
bandwidth $E_{W}=43$ meV for the highest moir\'e band in the MoTe${}_2$ layer. Therefore,
the band inversion between the highest moir\'e band in the MoTe${}_2$ layer and the
highest moir\'e band in the WSe${}_2$ layer can not be achieved with the out-of-plane
electric fields imposed in the experiment. In our theory, the band inversion is intrinsic.
The highest two moir\'e hole bands with opposite valley numbers in the MoTe${}_2$ layer
cross with each other, and the gap opening is attributed to the formation of an in-plane
$120^{\circ}$-N\'eel order and an in-plane ferromagnetic order. The N\'eel order ensures
the insulating gap. The Chern band emerges along with the formation of the ferromagnetic
order. Since the gap is opened before and after the topological phase transition, there is
no charge gap closure.
\item An in-plane ferromagnetic order emerges in the moir\'e superlattice under sufficient
out-of-plane electric fields due to exciton condensation. The exciton ferromagnetism can
be demonstrated by an EBH model and BKT transition. The ferromagnetic transition
temperature is estimated to be $5$ K, which is coincident with the observation.
\item The exciton Mott transition, a phase transition from exciton liquid to electron-hole
plasma, could occur as the electric-field strength reaches about $0.70$ V/nm. At low
temperatures, the exciton liquid becomes a BEC and the electron-hole plasma becomes a
BCS-like state known as an electron-hole condensate, which can be seen as a two-component
Fermi liquid. The continuous insulator-to-metal transition and the Fermi liquid behavior
at low temperatures could be explained by the excitonic BCS-BEC crossover.
\item The spin-polarized or valley-coherent QAH ground state across two layers can be
interpreted by interlayer-exciton condensation. Based on band-edge energies in
Eq.~(\ref{band_edge2}) and Fig.~\ref{fig:HF_band} (f), it is found that the hole-occupied
Chern band in the MoTe${}_2$ layer is mainly composed of the valley-polarized hole band
with $\tau=+$ as $M_z>0$ or mainly composed of the valley-polarized hole band with
$\tau=-$ as $M_z<0$. The exciton is formed by the vertical hole transition from the
MoTe${}_2$ layer to the WSe${}_2$ layer. By examining the band structures of the moir\'e
heterobilayers in Fig.~\ref{fig:moire_band}, the vertical transition from the
valley-polarized hole band generates a valley-coherent exciton, where the electron and the
hole reside in different valleys. Based on the spin-valley coupling shown in
Fig.~\ref{fig:moire_BZ} (c), the valley-coherent exciton is spin polarized. Additionally,
by the layer-selected Zeeman shifts shown in Eq.~(\ref{band_edge1}) and
Eq.~(\ref{band_edge2}), the spin-aligned MCD signal for exciton polarons in two layers can
be interpreted.
\item Full spin-valley polarization is not required for quantized Hall transport since the
QAH state is generated by the in-plane ferromagnetic order, not field-induced valley
polarization of holes. The observed canted spin texture can be explained by the
coexistence of the in-plane $120^{\circ}$-N\'eel order and the field-induced valley
polarization in the MoTe${}_2$ layer as the discussion in Sec.~\ref{sec:gap_opening}.
\item The QSH effect at $\nu=2$ hole filling and the band-to-QSH transition are not
studied in this work. This effect and this transition have been interpreted by Kane-Mele
physics\cite{exp2}. Our theory does not exclude the interpretation. It is worth noting
that the valence-band energy offset ($280$ meV) could be compensated by
Coulomb-interaction-driven band-energy renormalization, which contributes about $-110$ meV
energy shift. Since the bandwidth of the hole band in the MoTe${}_2$ layer has contributed
about $-40$ meV energy shift, the topological phase transition could occur at $-130$ meV
electric-field-induced energy shift ($0.50$ V/nm electric-field strength).
\end{enumerate}
Through these discussions, the consistency between the present theory and the experimental
observations is argued. Discussions about the Mott insulating state and exciton Mott
transition could also contribute to the study of the continuum phase transition found in
AA-stacked MoTe${}_2$/WSe${}_2$ heterobilayers.

An additional argument to support the present theory is the sparseness of QAH states being
found in transition metal dichalcogenide (TMDC) moir\'e heterobilayers. In fact, to the
best of the authors' knowledge, except AB-stacked MoTe${}_2$/WSe${}_2$ heterobilayers, no
QAH state has been found in other TMDC moir\'e heterobilayers. Several theories that
explain the QAH effect in AB-stacked MoTe${}_2$/WSe${}_2$ heterobilayers could predict a
wide distribution of QAH states in TMDC moir\'e heterobilayers. Nevertheless, it seems to
be not the case. In our theory, the sparseness can be attributed to the restricted
parametrization for the present model to meet the conditions that exciton ferromagnetism
occurs and the ferromagnetic phase transition precedes the exciton Mott transition.

In conclusion, a theory to explain the QAH effect and the topological phase transition in
AB-stacked MoTe${}_2$/WSe${}_2$ heterobilayers is provided. The consistency between the
theory and experimental observations is argued. This work may contribute a new viewpoint
to search QAH insulators among correlated materials and a new route to study topological
orders in moir\'e materials.

\begin{acknowledgments}

This work was supported in part by the National Science and Technology Council, Taiwan
(Contract Nos. 109-2112-M-001-046 and 110-2112-M-001-042), the Ministry of Education,
Taiwan (Higher Education Sprout Project NTU-111L104022), and the National Center for
Theoretical Sciences of Taiwan. We thank reviewers for asking critical questions and
providing valuable comments on the preliminary version of this work.

\end{acknowledgments}

\appendix

\section{Moir\'e band-structure calculation\label{sec:moire_band}}

The method to calculate moir\'e band structures for the continuum model of AB-stacked
$\text{MoTe}_2$/$\text{WSe}_2$ heterobilayers is given in this section. In
Sec.~\ref{sec:plane_wave}, the plane-wave method to solve moir\'e band structures is
introduced. In Sec.~\ref{sec:HF_band}, the Hartree-Fock approximation for band structure
calculation is reviewed.

\subsection{Plane-wave method\label{sec:plane_wave}}

The single-particle wavefunction of a carrier in the moir\'e superlattice can be expanded
in terms of plane-wave basis functions as
\begin{eqnarray}
  \psi_{n\mathbf{k}}(\mathbf{r})
  =
  \sum_{{\alpha}}u_{n\mathbf{k}}
  \phi_{\mathbf{G}_{\alpha},\mathbf{k}}(\mathbf{r}),
\end{eqnarray}
where $\phi_{\mathbf{G}_{\alpha},\mathbf{k}}(\mathbf{r})$ denotes a plane-wave basis
function and $u_{{\alpha},n\mathbf{k}}$ is the the expansion coefficient. The plane-wave
basis function is written as
\begin{eqnarray}
  \phi_{\mathbf{G},\mathbf{k}}(\mathbf{r})
  =
  e^{\mathtt{i}\left(\mathbf{k}-\mathbf{G}\right)\cdot\mathbf{r}}/\sqrt{S},
\end{eqnarray}
with $S=N(\sqrt{3}/2)a^2_{\text{M}}$ the area of the moir\'e lattice. By using the
plane-wave expansion, the Hamiltonian matrix is given by $\langle
a,\mathbf{G}_1|\tilde{H}_{\mathbf{k}}|b,\mathbf{G}_2\rangle =
\int\phi^*_{\mathbf{G}_1,\mathbf{k}}(\mathbf{r})H_{ab}(\mathbf{r})
\phi_{\mathbf{G}_2,\mathbf{k}}(\mathbf{r})\text{d}^2r$ and the diagonal part is given by
\begin{eqnarray}
  \langle a,\mathbf{G}_\alpha|\tilde{H}_{\mathbf{k}}|a,\mathbf{G}_\beta\rangle
  &=&
  \delta_{\alpha\beta}
  \frac{|\mathbf{k}-\mathbf{G}_{\alpha}-\tau_{a}\bsym{\kappa}_{l_a}|^2}{2m_{l_a}}\n
  &&-\langle a,\mathbf{G}_\alpha|\tilde{V}_{l_a,\mathbf{k}}|a,\mathbf{G}_\beta\rangle,
\end{eqnarray}
with $\langle a,\mathbf{G}_\alpha|\tilde{V}_{l_a,\mathbf{k}}|a,\mathbf{G}_\beta\rangle =
-\mathtt{i}V \sum_{j} (-1)^{l_a+j}\delta(\mathbf{G}_{\alpha}
-\mathbf{G}_{\beta}-\bsym{g}_{j})$. The off-diagonal Hamiltonian matrix element is given
by
\begin{eqnarray}
  \langle a,\mathbf{G}_\alpha|\tilde{H}_{\mathbf{k}}|b,\mathbf{G}_\beta\rangle
  &=&
  w\delta_{\tau_a,\tau_b}\delta_{l_a,l_b+1}
  \Big[\delta(\mathbf{G}_\alpha-\mathbf{G}_\beta)\n
  &&+\sum_{j=1,2}\delta(\mathbf{G}_\alpha-\mathbf{G}_\beta-\tau_a\bsym{g}_{j})\Big]\n
  &&+w\delta_{\tau_a,\tau_b}\delta_{l_a+1,l_b}
  \Big[\delta(\mathbf{G}_\alpha-\mathbf{G}_\beta)\n
  &&+\sum_{j=1,2}\delta(\mathbf{G}_\alpha-\mathbf{G}_\beta+\tau_a\bsym{g}_{j})\Big].
  \hskip4ex
\end{eqnarray}

\subsection{Hartree-Fock approximation\label{sec:HF_band}}

Given the many-particle Hamiltonian for multi-component particle fields in
Eq.~(\ref{many_particle}), the quasiparticle creation and annihilation operators can be
transformed as $\hat{\Psi}^{\dagger}_{a}(\mathbf{r}) =
\sum_{n\mathbf{k}}\psi^*_{a,n\mathbf{k}}(\mathbf{r})\hat{d}^{\dagger}_{n\mathbf{k}}$,
$\hat{\Psi}_{a}(\mathbf{r}) =
\sum_{n\mathbf{k}}\psi_{a,n\mathbf{k}}(\mathbf{r})\hat{d}_{n\mathbf{k}}$, with
$\psi_{a,n\mathbf{k}}(\mathbf{r})$ the quasiparticle wavefunction. By using the
variational method, it is found that the quasiparticle wavefunction can be solved by the
Hartree-Fock equation
\begin{eqnarray}
  \sum_{b}\int F_{ab}(\mathbf{r}_{1},\mathbf{r}_{2})
  \psi_{b,n\mathbf{k}}(\mathbf{r}_2)\text{d}^2r_2
  =
  \varepsilon_{n\mathbf{k}}\psi_{a,n\mathbf{k}}(\mathbf{r}_1),\hskip2ex
\end{eqnarray}
where $n$ is the band index, $\varepsilon_{n\mathbf{k}}$ is the quasiparticle energy. The
Fock operator is defined by
\begin{eqnarray}
  F_{ab}(\mathbf{r}_{1},\mathbf{r}_{2})
  &=&
  \delta(\mathbf{r}_{1}-\mathbf{r}_{2})\left[H_{ab}(\mathbf{r}_1)
  +\delta_{ab}K_{aa}(\mathbf{r}_1)\right]\n
  &&-J_{ab}(\mathbf{r}_{1},\mathbf{r}_{2}),
\end{eqnarray}
where $K_{aa}(\mathbf{r}_1) \equiv
\sum_{c}\int{W}_{ac}(\mathbf{r}_{13})\rho_{cc}(\mathbf{r}_3,\mathbf{r}_3)\text{d}^2r_3$ is
the Coulomb operator, $J_{ab}(\mathbf{r}_{1},\mathbf{r}_{2}) \equiv
W_{ab}(\mathbf{r}_{12})\rho_{ba}(\mathbf{r}_{2},\mathbf{r}_{1})$ is the exchange operator,
and
\begin{eqnarray}
  \rho_{ab}(\mathbf{r}_{1},\mathbf{r}_{2})
  &=&
  \sum_{n\mathbf{k}}n_{n\mathbf{k}}
  \psi_{a,n\mathbf{k}}(\mathbf{r}_1)\psi^*_{b,n\mathbf{k}}(\mathbf{r}_2)
\end{eqnarray}
is the density-matrix operator, where
$n_{n\mathbf{k}}=\langle{\hat{d}^{\dagger}_{n\mathbf{k}}\hat{d}_{n\mathbf{k}}}\rangle$ is
the occupation number of charges in the band $n$ with the momentum $\mathbf{k}$.

By using the plane-wave method, the wavefunction coefficient can be obtained by solving
the Hartree-Fock equation $\tilde{F}_{\mathbf{k}}u_{n\mathbf{k}}
=\varepsilon_{n\mathbf{k}}u_{n\mathbf{k}}$. The Fock matrix is given by
\begin{eqnarray}
  \langle a,\mathbf{G}_\alpha|\tilde{F}_{\mathbf{k}}|b,\mathbf{G}_\beta\rangle
  &=&
  \langle a,\mathbf{G}_\alpha|\tilde{H}_{\mathbf{k}}|b,\mathbf{G}_\beta\rangle
  -\langle a,\mathbf{G}_\alpha|\tilde{J}_{\mathbf{k}}|b,\mathbf{G}_\beta\rangle\n
  &&+\delta_{ab}\langle a,\mathbf{G}_\alpha|
  \tilde{K}_{\mathbf{k}}|a,\mathbf{G}_\beta\rangle.
\end{eqnarray}
The Coulomb integral is given by
\begin{eqnarray}
  \langle a,\mathbf{G}_\alpha|\tilde{K}_{\mathbf{k}}|a,\mathbf{G}_\beta\rangle
  &=&
  \sum_{\mathbf{q},\mathbf{G},c}
  \langle c,\mathbf{G}+\mathbf{G}_{\alpha}|\tilde{P}_{\mathbf{q}}
  |c,\mathbf{G}+\mathbf{G}_{\beta}\rangle\n
  &&\times{\tilde{W}_{ac}(\mathbf{G}_{\alpha}-\mathbf{G}_{\beta})}/S,
\end{eqnarray}
and the exchange integral is given by
\begin{eqnarray}
  \langle a,\mathbf{G}_\alpha|\tilde{J}_{\mathbf{k}}|b,\mathbf{G}_\beta\rangle
  &=&
  \sum_{\mathbf{q},\mathbf{G}}
  \langle a,\mathbf{G}+\mathbf{G}_{\alpha}|\tilde{P}_{\mathbf{q}}
  |b,\mathbf{G}+\mathbf{G}_{\beta}\rangle\n
  &&\times\tilde{W}_{ab}(\mathbf{G}+\mathbf{k}-\mathbf{q})/S,
\end{eqnarray}
where
\begin{eqnarray}
  \langle a,\mathbf{G}_\alpha|\tilde{P}_{\mathbf{k}}|b,\mathbf{G}_\beta\rangle
  =
  \sum_{n}n_{n\mathbf{k}}
  u_{(a,\mathbf{G}_\alpha),n\mathbf{k}}u^*_{(b,\mathbf{G}_\beta),n\mathbf{k}}\hskip4ex
\end{eqnarray}
is the single-particle projection matrix and $\tilde{W}_{ab}(\mathbf{k}) =\int
\exp\left({-\mathtt{i}\mathbf{k}\cdot\mathbf{r}}\right) W_{ab}(\mathbf{r})\text{d}^2r$ is
the screened Coulomb potential.

\section{Formulations for exciton condensation\label{sec:exciton_condensation}}

In this section, theory of exciton condensation is revisited based on
references\cite{jerome1967excitonic, zittartz1967theory, halperin1968excitonic,
keldysh1968collective, comte1982exciton, nozieres1982exciton, fernandez1997spin,
chu1996theory, wu2015theory}. In Sec.~\ref{sec:variation}, the gap equation for exciton
condensates is derived. In Sec.~\ref{sec:exciton_binding_energy}, the variational method
to solve the exciton binding energy is introduced. In
Sec.~\ref{sec:excitonic_instability}, the conditions of excitonic instability are
discussed.

\subsection{Gap equation\label{sec:variation}}

Exciton condensation can be described by the EHS Hamiltonian in Eq.~(\ref{eh_Hamiltonian})
with a more general form of the combination of electron-band and hole-band dispersion
\begin{eqnarray}
  \varepsilon^{\text{e}}_{-\mathbf{k}}+\varepsilon^{\text{h}}_{\tau,\mathbf{k}}
  =
  \tilde{D}+\frac{k^2}{2\mu_{\text{X}}},
  \label{gap_eq_para1}
\end{eqnarray}
where $\mu_{\text{X}}=m_{\text{e}}m_{\text{h}}/(m_{\text{e}}+m_{\text{h}})$ is the reduced
mass. The variational state for a exciton condensate is assumed to be the following BCS
state $|{\Phi_{\text{BCS}}}\rangle = \prod_{\tau,\mathbf{k}}
\left(u_{\tau,\mathbf{k}}+v_{\tau,\mathbf{k}}
\hat{d}^{\dagger}_{\tau,\mathbf{k}}\hat{c}^{\dagger}_{-\mathbf{k}}\right)
|{\Phi_\text{HF}}\rangle$, where $u_{\tau,\mathbf{k}}$ and $v_{\tau,\mathbf{k}}$ are
variational coefficients subject to the normalization condition
$u_{\tau,\mathbf{k}}^2+v_{\tau,\mathbf{k}}^2=1$. Note that
\begin{eqnarray}
  \frac{\delta u_{\tau,\mathbf{k}}}{\delta v_{\tau',\mathbf{k}'}}
  =
  -\delta_{\tau,\tau'}\delta_{\mathbf{k},\mathbf{k}'}
  \frac{v_{\tau,\mathbf{k}}}{u_{\tau,\mathbf{k}}}.
\end{eqnarray}
The expectation of the Hamiltonian is given by
\begin{eqnarray}
  \langle{\hat{\mathcal{H}}_{\text{EHS}}}\rangle
  &=&
  \sum_{\tau,\mathbf{k}}
  \left(\varepsilon^{\text{e}}_{-\mathbf{k}}
  +\varepsilon^{\text{h}}_{\tau,\mathbf{k}}\right)v_{\tau,\mathbf{k}}^2\n
  &&-\sum_{\tau,\mathbf{k}\neq\mathbf{k}'}
  \frac{W^{\text{eh}}_{\mathbf{k}'-\mathbf{k}}}{S}
  v_{\tau,\mathbf{k}'}v_{\tau,\mathbf{k}}
  u_{\tau,\mathbf{k}}u_{\tau,\mathbf{k}'}.
\end{eqnarray}
The variation of the energy expectation value is given by
\begin{eqnarray}
  \frac{\delta\langle{\hat{\mathcal{H}}_{\text{EHS}}}\rangle}{\delta v_{\tau,\mathbf{k}}}
  &=&
  2\left(\varepsilon^{\text{e}}_{-\mathbf{k}}+\varepsilon^{\text{h}}_{\tau,\mathbf{k}}
  \right)v_{\tau,\mathbf{k}}\n
  &&-2\sum_{\mathbf{k}'}\frac{W^{\text{eh}}_{\mathbf{k}-\mathbf{k}'}}{S}
  v_{\tau,\mathbf{k}'}u_{\tau,\mathbf{k}'}\left(u_{\tau,\mathbf{k}}
  -\frac{v^2_{\tau,\mathbf{k}}}{u_{\tau,\mathbf{k}}}\right).\n
\end{eqnarray}
By variation ${\delta\langle{\hat{\mathcal{H}}_{\text{EHS}}}\rangle}/{\delta
v_{\tau,\mathbf{k}}}=0$, and by assuming $\Delta_{\tau,\mathbf{k}} =
(1/S)\sum_{\mathbf{k}'}W^{\text{eh}}_{\mathbf{k}-\mathbf{k}'}
v_{\tau,\mathbf{k}'}u_{\tau,\mathbf{k}'}$, ${\Xi}_{\tau,\mathbf{k}} =
\big(\varepsilon^{\text{e}}_{-\mathbf{k}}+\varepsilon^{\text{h}}_{\tau,\mathbf{k}}
\big)/2$, we get $\big(2v_{\tau,\mathbf{k}}u_{\tau,\mathbf{k}}\big){\Xi}_{\tau,\mathbf{k}}
-\big(u^2_{\tau,\mathbf{k}}-v^2_{\tau,\mathbf{k}}\big)\Delta_{\tau,\mathbf{k}}=0$. By
replacing $u_{\tau,\mathbf{k}}=\cos\theta_{\tau,\mathbf{k}}$ and
$v_{\tau,\mathbf{k}}=\sin\theta_{\tau,\mathbf{k}}$, we find
$\sin(2\theta_{\tau,\mathbf{k}}){\Xi}_{\tau,\mathbf{k}} =
\cos(2\theta_{\tau,\mathbf{k}})\Delta_{\tau,\mathbf{k}}$ and
$\cos(2\theta_{\tau,\mathbf{k}})={\Xi}_{\tau,\mathbf{k}}/\mathcal{E}_{\tau,\mathbf{k}}$,
with $\mathcal{E}_{\tau,\mathbf{k}} =
\sqrt{|{\Xi}_{\tau,\mathbf{k}}|^2+|\Delta_{\tau,\mathbf{k}}|^2}$. Therefore, we get
\begin{eqnarray}
  u^2_{\tau,\mathbf{k}}
  =
  \frac{1}{2}\left(1+\frac{{\Xi}_{\tau,\mathbf{k}}}{\mathcal{E}_{\tau,\mathbf{k}}}\right),
  \hskip1ex
  v^2_{\tau,\mathbf{k}}
  =
  \frac{1}{2}\left(1-\frac{{\Xi}_{\tau,\mathbf{k}}}{\mathcal{E}_{\tau,\mathbf{k}}}\right),
\end{eqnarray}
and $2v_{\tau,\mathbf{k}}u_{\tau,\mathbf{k}} = \sin(2\theta_{\tau,\mathbf{k}}) =
\Delta_{\tau,\mathbf{k}}/\mathcal{E}_{\tau,\mathbf{k}}$. The gap equation can be found as
Eq.~(\ref{gap_eq_1}). By replacing the variational parameters, the exciton density is
given by
\begin{eqnarray}
  {n}_{\text{X}}
  =
  \frac{1}{N}\sum_{\tau,\mathbf{k}}v^2_{{\tau},\mathbf{k}}
  =
  \sum_{\tau,\mathbf{k}}
  \frac{1}{2N}\left(1-\frac{{\Xi}_{\tau,\mathbf{k}}}{\mathcal{E}_{\tau,\mathbf{k}}}\right).
  \label{exciton_density}
\end{eqnarray}

\subsection{Exciton binding energy\label{sec:exciton_binding_energy}}

The variational method to solve the exciton banding energy is introduced in this section.
Two-dimensional Slater-type orbitals (STOs) are used to expanded the variational exciton
wavefunction. A more detailed discussion of this method can be found in
Ref.~\cite{mypaper0}. As the combination of electron and hole kinetic energies is assumed
to be given by Eq.~(\ref{gap_eq_para1}), the interlayer exciton wavefunction
$\Psi_{I}(\mathbf{r})$ can be solved by the Schr\"{o}dinger equation
\begin{eqnarray}
  \left[\tilde{D}-\frac{\nabla^2}{2\mu_{\text{X}}}-W(\mathbf{r})\right]
  \Psi_{I}(\mathbf{r})
  =
  E_{I}\Psi_{I}(\mathbf{r}).
  \label{exciton_eq0}
\end{eqnarray}
The Fourier transform of the exciton wavefunction can be found by
$\tilde{\Psi}_{I}({\mathbf{k}}) = \int\;e^{-\mathtt{i}\mathbf{k}\cdot\mathbf{r}}
\Psi_{I}(\mathbf{r})\text{d}^2r$. The exciton wavefunction can be expanded as
\begin{eqnarray}
  \tilde{\Psi}_{I}({\mathbf{k}})
  =
  \sum_{\alpha}\mathcal{U}_{\alpha,I}\tilde{\Phi}_{\alpha}(\mathbf{k}),
  \label{exciton_wavefunction}
\end{eqnarray}
where $\mathcal{U}_{\alpha,I}$ is the wavefunction coefficient and
$\tilde{\Phi}_{\alpha}(\mathbf{k})$ is the basis function. The exciton wavefunction
coefficient can be solved by the eigenvalue equation
\begin{eqnarray}
  \sum_{\beta}
  \left(\mathcal{T}_{\alpha\beta}-\mathcal{W}_{\alpha\beta}\right)\mathcal{U}_{\beta,I}
  =
  E_{I}\sum_{\beta}\mathcal{O}_{\alpha\beta}\mathcal{U}_{\beta,I},
  \label{exciton_eq}
\end{eqnarray}
where $\mathcal{T}_{\alpha\beta} \equiv -[1/({2\mu_{\text{X}}})]
\int\Phi^*_{\alpha}(\mathbf{r})\nabla^2\Phi_{\beta}(\mathbf{r})\text{d}^2{r}$,
$\mathcal{W}_{\alpha\beta} \equiv
\int{\Phi}^*_{\alpha}(\mathbf{r}){W}(\mathbf{r}){\Phi}_{\beta}(\mathbf{r})\text{d}^2r$,
$\mathcal{O}_{\alpha\beta} \equiv
\int\Phi^*_{\alpha}(\mathbf{r})\Phi_{\beta}(\mathbf{r})\text{d}^2{r}$ are the kinetic
integral, the potential integral, and the overlap integral. By using an orthogonalization
transformation $\bar{\mathcal{U}}_{\alpha,I}=
\sum_{\beta}\mathcal{O}^{1/2}_{\alpha\beta}\mathcal{U}_{\beta,I}$, the eigenvalue equation
becomes
\begin{eqnarray}
  \sum_{\beta}\Omega_{\alpha\beta}
  \bar{\mathcal{U}}_{\beta,I}
  =
  E_{I}\bar{\mathcal{U}}_{\beta,I},
  \label{exciton_eq2}
\end{eqnarray}
with $\Omega_{\alpha\beta} = \sum_{\alpha'\beta'} \mathcal{O}^{-1/2}_{\alpha\alpha'}
\left(\mathcal{T}_{\alpha'\beta'}-\mathcal{W}_{\alpha'\beta'}\right)
\mathcal{O}^{-1/2}_{\beta'\beta}$.

To solve the exciton eigenvalue equation in Eq.~(\ref{exciton_eq}), the basis function can
be given by a two-dimensional STO, which is written as
\begin{eqnarray}
  \Phi_{\alpha}(\mathbf{r})
  &=&
  \frac{e^{\mathtt{i}L_{\alpha}\varphi}}{\sqrt{2\pi}}r^{N_{\alpha}-1}
  e^{-\mathcal{Z}_{\alpha} r},
  \label{sto}
\end{eqnarray}
where $N_{\alpha}$, $L_{\alpha}$ are the principal quantum number and angular-momentum
quantum number of the orbital $\Phi_{\alpha}$, $\mathcal{Z}_{\alpha}$ is the shielding
constant, and $\varphi$ is the azimuth angle. Several different values of
$\mathcal{Z}_{\alpha}$ can be used to find the optimum shape of the radial part of the
wavefunction. The Fourier transform of the two-dimensional STO can be written as
\begin{eqnarray}
  \tilde{\Phi}_{\alpha}(\mathbf{k})
  &=&
  \int\Phi_{\alpha}(\mathbf{r})e^{-\mathtt{i}\mathbf{k}\cdot\mathbf{r}}\text{d}^2r
  =
  \frac{e^{\mathtt{i}L_{\alpha}\varphi_{\mathbf{k}}}}{\sqrt{2\pi}}
  \tilde{\mathcal{R}}_{N_{\alpha},L_{\alpha}}(\mathcal{Z}_{\alpha},k),\n
\end{eqnarray}
where the radial function in momentum space can be obtained by the generating formula
\begin{eqnarray}
  \tilde{\mathcal{R}}_{N,L}(\mathcal{Z},k)
  &=&
  \frac{2\pi(-\mathtt{i})^{N}}{k^{N+1}}
  \frac{\text{d}^N}{\text{d}z^N}
  \frac{\left(z-\mathtt{i}\eta\sqrt{1-z^2}\right)^{|L|}}{\sqrt{1-z^2}}
  \Bigg\vert_{z=\mathtt{i}{\mathcal{Z}}/{k}},\n
  \label{Rk_formula}
\end{eqnarray}
with $\eta=L/|L|$. The kinetic integral is given by
\begin{eqnarray}
  \mathcal{T}_{\alpha\beta}
  &=&
  -\frac{\delta_{L_\alpha,L_\beta}}{2\mu_{\text{X}}}
  \frac{(N_\alpha+N_\beta-1)!}
  {\left(\mathcal{Z}_\alpha+\mathcal{Z}_\beta\right)^{N_\alpha+N_\beta}}
  \Bigg\{(1-\delta_{N_\beta,1})\n
  &&\times\frac{\big[(N_\beta-1)^2-L_\beta^2\big]
  \left(\mathcal{Z}_\alpha+\mathcal{Z}_\beta\right)^2}
  {(N_\alpha+N_\beta-1)(N_\alpha+N_\beta-2)}\n
  &&-\frac{\left[(2N_\beta-1)\mathcal{Z}_\beta\right]
  \left(\mathcal{Z}_\alpha+\mathcal{Z}_\beta\right)}
  {(N_\alpha+N_\beta-1)}+\mathcal{Z}^2_\beta\Bigg\}.
\end{eqnarray}
The overlap integral is given by
\begin{eqnarray}
  \mathcal{O}_{\alpha\beta}
  &=&
  \delta_{L_\alpha,L_\beta}\frac{(N_\alpha+N_\beta-1)!}
  {(\mathcal{Z}_\alpha+\mathcal{Z}_\beta)^{N_\alpha+N_\beta}}.
\end{eqnarray}
The potential integral is given by
\begin{eqnarray}
  \mathcal{W}_{\alpha\beta}
  &=&
  \frac{\delta_{L_{\alpha},L_{\beta}}}{(2\pi)^2}\int^{\infty}_0
  \tilde{\mathcal{R}}_{N_{\alpha}+N_{\beta}-1,0}
  (\mathcal{Z}_{\alpha}+\mathcal{Z}_{\beta},k)\tilde{W}(k)k\text{d}k.\n
  \label{potential_integral}
\end{eqnarray}
The Coulomb potential can be given by $\tilde{W}(k)=W^{\text{eh}}_{\mathbf{k}}$. An
exciton $I=(N,L)$ can be indicated by a principal quantum number $N$ and an angular
momentum $L$, with $L$ being a constant for every orbital in the exciton wavefunction. In
the present study, only $L=0$ is considered.

\subsection{Conditions of excitonic instability\label{sec:excitonic_instability}}

If the number of excitons is restricted to one, by a variation of the ground-state
expectation value of the EHS Hamiltonian
\begin{eqnarray}
  \delta\left[\langle{\Phi|\hat{\mathcal{H}}_{\text{EHS}}|\Phi}\rangle
  -\lambda\left(N{n}_{\text{X}}-1\right)\right]
  /\delta{v}_{\tau,\mathbf{k}}=0,
\end{eqnarray}
with $\lambda$ the Lagrange multiplier, the BCS variational coefficient
$v_{{\tau},\mathbf{k}}$ can be solved by
\begin{eqnarray}
  [k^2/(2\mu_{\text{X}})+E_{\text{X}}]v_{{\tau},\mathbf{k}}
  =
  \sum\nolimits_{\mathbf{k}'}W^{\text{eh}}_{\mathbf{k}-\mathbf{k}'}v_{{\tau},\mathbf{k}'}/S,
  \label{exciton_eq1}
\end{eqnarray}
where $E_{\text{X}}=\tilde{D}-\lambda$ is the exciton binding energy. By comparing
Eq.~\ref{exciton_eq1} with the exciton equation in Eq.~(\ref{exciton_eq0}), the
coefficient is equivalent to the ground-state exciton wavefunction
($v_{{\tau},\mathbf{k}}=\Psi_{I,\mathbf{k}}$ for $I=0$) under the condition. We define the
excitonic instability to form condensation by the condition
\begin{eqnarray}
  \Delta_{\tau,\mathbf{k}}\neq{0}
\end{eqnarray}
for at least one $(\tau$, $\mathbf{k})$ state. In this section, we want to show that the
necessary condition for the excitonic instability at zero temperature is $\tilde{D}\leq
E_{\text{X}}$ and the sufficient condition is $\tilde{D}<E_{\text{X}}$, with $\tilde{D}$
being the effective band gap.

To prove the necessary condition, we can rewrite the gap equation by defining
\begin{eqnarray}
  \gamma_{\tau,\mathbf{k}}
  \equiv
  u_{\tau,\mathbf{k}}v_{\tau,\mathbf{k}}
  =
  \frac{\Delta_{\tau,\mathbf{k}}}{2\mathcal{E}_{\tau,\mathbf{k}}}.
  \label{gap_eq_para2}
\end{eqnarray}
The gap equation can be rewritten as $\gamma_{\tau,\mathbf{k}} =
[{1}/({2\mathcal{E}_{\tau,\mathbf{k}}})]
\sum_{\mathbf{k}'}{W^{\text{eh}}_{\mathbf{k}-\mathbf{k}'}} \gamma_{\tau,\mathbf{k}'}/S$,
and then it can become
\begin{eqnarray}
  2\mathcal{E}_{\tau,\mathbf{k}}\gamma_{\tau,\mathbf{k}}
  -\sum_{\mathbf{k}'}\frac{W^{\text{eh}}_{\mathbf{k}-\mathbf{k}'}}{S}\gamma_{\tau,\mathbf{k}'}
  =
  0.
\end{eqnarray}
By using
\begin{eqnarray}
  \mathcal{E}_{\tau,\mathbf{k}}
  &=&
  \sqrt{|{\Xi}_{\tau,\mathbf{k}}|^2+|\Delta_{\tau,\mathbf{k}}|^2}
  =
  |{\Xi}_{\tau,\mathbf{k}}|
  \sqrt{1+|\Delta_{\tau,\mathbf{k}}|^2/|{\Xi}_{\tau,\mathbf{k}}|^2}\n
  &=&
  |{\Xi}_{\tau,\mathbf{k}}|+g_{\tau,\mathbf{k}},
\end{eqnarray}
with $g_{\tau,\mathbf{k}} =
|{\Xi}_{\tau,\mathbf{k}}|\big(\sqrt{1+{|\Delta_{\tau,\mathbf{k}}|^2}/
{|{\Xi}_{\tau,\mathbf{k}}|^2}}-1\big)\geq{0}$, and assuming ${\Xi}_{\tau,\mathbf{k}}>0$,
the gap equation becomes
\begin{eqnarray}
  2({\Xi}_{\tau,\mathbf{k}}+g_{\tau,\mathbf{k}})\gamma_{\tau,\mathbf{k}}
  -\sum_{\mathbf{k}'}\frac{W^{\text{eh}}_{\mathbf{k}-\mathbf{k}'}}{S}\gamma_{\tau,\mathbf{k}'}
  =
  0.
  \label{gap_eq_3}
\end{eqnarray}
By using Eq.~(\ref{gap_eq_para1}), the gap equation can be rewritten as
\begin{eqnarray}
  \sum_{\mathbf{k}'}\left(\mathcal{A}_{\mathbf{k},\mathbf{k}'}
  +\mathcal{B}_{\mathbf{k},\mathbf{k}'}\right)
  \gamma_{\tau,\mathbf{k}'}
  =
  0,
\end{eqnarray}
with $\mathcal{A}_{\mathbf{k},\mathbf{k}'}
=
\delta_{\mathbf{k},\mathbf{k}'}
\big[\tilde{D}+k^2/(2\mu_{\text{X}})-{W^{\text{eh}}_{\mathbf{k}-\mathbf{k}'}}/{S}\big]$
and $\mathcal{B}_{\mathbf{k},\mathbf{k}'} =
\delta_{\mathbf{k},\mathbf{k}'}2g_{\tau,\mathbf{k}}$.
Note that $g_{\tau,\mathbf{k}}$ and $\gamma_{\tau,\mathbf{k}}$ become independent of
$\tau$ because Eq.~(\ref{gap_eq_para1}) is used. A trivial solution
($\gamma_{\tau,\mathbf{k}}=0$) of the equation leads to $\Delta_{\tau,\mathbf{k}}=0$ for
every $(\tau,\mathbf{k})$ state. The equation has nontrivial solutions of
$\gamma_{\tau,\mathbf{k}}$ only if $\textit{Det}\left(\mathcal{A}_{\mathbf{k},\mathbf{k}'}
+\mathcal{B}_{\mathbf{k},\mathbf{k}'}\right)=0$, which implies the existence of at least a
zero eigenvalue for matrix $\bsym{\mathcal{A}}+\bsym{\mathcal{B}}$. Since
$g_{\tau,\mathbf{k}}\geq{0}$ for each $\tau$ and $\mathbf{k}$, matrix $\bsym{\mathcal{B}}$
is positive semi-definite. If matrix $\bsym{\mathcal{A}}$ is positive definite, matrix
$\bsym{\mathcal{A}}+\bsym{\mathcal{B}}$ will be positive definite, which contradicts to
that matrix $\bsym{\mathcal{A}}+\bsym{\mathcal{B}}$ has at least a zero eigenvalue.
Therefore, matrix $\bsym{\mathcal{A}}$ is not positive definite. It indicates the lowest
eigenvalue of matrix $\bsym{\mathcal{A}}$ is not a positive number. By using the exciton
equation in Eq.~(\ref{exciton_eq1}), the lowest eigenvalue of matrix $\bsym{\mathcal{A}}$
is solved by
\begin{eqnarray}
  \sum_{\mathbf{k}'}\mathcal{A}_{\mathbf{k},\mathbf{k}'}\Psi_{I,\mathbf{k}'}
  =
  E_{I}\Psi_{I,\mathbf{k}},
  \label{excitation_equation}
\end{eqnarray}
with $E_{I}$ the eigenvalue and $\Psi_{I,\mathbf{k}}$ the eigenfunction. The lowest
eigenvalue of the equation is given by $E_{I=0}=\tilde{D}-E_{\text{X}}$. Therefore, the
condition for matrix $\bsym{\mathcal{A}}$ being not positive is given by
$E_{0}=\tilde{D}-E_{\text{X}}\leq{0}$, which gives the necessary condition $\tilde{D}\leq
E_{\text{X}}$ for excitonic instability.

To prove the sufficient condition, we assume that the eigenvalues $E_I$ and eigenfunctions
$\Psi_{I,\mathbf{k}}$ of matrix $\bsym{\mathcal{A}}$ are given by
Eq.~(\ref{excitation_equation}), and the parameter $\gamma_{\tau,\mathbf{k}}$ can be
expanded by the eigenfunctions
\begin{eqnarray}
  \gamma_{\tau,\mathbf{k}}
  =
  \sum_{I}C_{I}\Psi_{I,\mathbf{k}},
\end{eqnarray}
with $C_{I}$ being variational coefficient. The gap equation can be reformulated as
\begin{eqnarray}
  \Delta_{\tau,\mathbf{k}}
  &=&
  \sum_{\mathbf{k}'}{W^{\text{eh}}_{\mathbf{k}-\mathbf{k}'}}\gamma_{\tau,\mathbf{k}'}/S\n
  &=&
  -\sum_{\mathbf{k}'}\left(\mathcal{A}_{\mathbf{k},\mathbf{k}'}
  -\delta_{\mathbf{k},\mathbf{k}'}2{\Xi}_{\tau,\mathbf{k}}
  \right)\sum_{I}C_{I}\Psi_{I,\mathbf{k}'}\n
  &=&
  \sum_{I}C_{I}\left(2{\Xi}_{\tau,\mathbf{k}}-E_{I}\right)\Psi_{I,\mathbf{k}}.
\end{eqnarray}
The gap equation becomes
\begin{eqnarray}
  \sum_{I}C_{I}\Psi_{I,\mathbf{k}}
  &=&
  \frac{\sum_{I}C_{I}\left(2{\Xi}_{\tau,\mathbf{k}}-E_{I}\right)
  \Psi_{I,\mathbf{k}}}{2\sqrt{{\Xi}^2_{\tau,\mathbf{k}}
  +\left[\sum_{I}C_{I}\left(2{\Xi}_{\tau,\mathbf{k}}
  -E_{I}\right)\Psi_{I,\mathbf{k}}\right]^2}}.\n
\end{eqnarray}
By assigning $\mathbf{k}=\mathbf{0}$, the gap equation becomes
\begin{eqnarray}
  \sum_{I}C_{I}\Psi_{I,\mathbf{0}}
  &=&
  \frac{\sum_{I}C_{I}\left(2{\Xi}_{\tau,\mathbf{0}}-E_{I}\right)
  \Psi_{I,\mathbf{0}}}{2\sqrt{{\Xi}^2_{\tau,\mathbf{0}}
  +\left[\sum_{I}C_{I}\left(2{\Xi}_{\tau,\mathbf{0}}
  -E_{I}\right)\Psi_{I,\mathbf{0}}\right]^2}}.\n
\end{eqnarray}
A good approximation for the exciton wavefunctions $\Psi_{I,\mathbf{k}}$ is to use the
wavefunctions solved from the two-dimensional hydrogen-atom problem. The wavefunctions of
two-dimensional hydrogen atoms have the properties $\Psi_{I,\mathbf{k}}={0}$ for $I>{0}$
and $\mathbf{k}=\mathbf{0}$. By assuming $\Psi_{I,\mathbf{0}}\simeq{0}$ for $I>{0}$, the
gap equation can be reduced to
\begin{eqnarray}
  1\simeq
  \frac{2{\Xi}_{\tau,\mathbf{0}}-E_{0}}{2\sqrt{{\Xi}^2_{\tau,\mathbf{0}}
  +\left[C_{0}\left(2{\Xi}_{\tau,\mathbf{0}}
  -E_{0}\right)\Psi_{0,\mathbf{0}}\right]^2}},
\end{eqnarray}
which leads to $\left(2C_{0}\Psi_{0,\mathbf{0}}\right)^2 \simeq
1-{4{\Xi}^2_{\tau,\mathbf{0}}}/{\left(2{\Xi}_{\tau,\mathbf{0}}-E_{0}\right)^2}$. If
$E_{0}<0$, there is a nontrivial solution for $|C_{0}|$, which is given by
\begin{eqnarray}
  |C_{0}|
  &\simeq&
  \frac{1}{2\Psi_{0,\mathbf{0}}}
  \sqrt{1-\left(\frac{2{\Xi}_{\tau,\mathbf{0}}}{2{\Xi}_{\tau,\mathbf{0}}
  -E_{0}}\right)^2}.
\end{eqnarray}
The gap order parameter can be given approximately by
\begin{eqnarray}
  \Delta_{\tau,\mathbf{k}}
  &\simeq&
  C_{0}\left(2{\Xi}_{\tau,\mathbf{k}}-E_{0}\right)\Psi_{0,\mathbf{k}}\n
  &=&
  \pm
  \frac{2{\Xi}_{\tau,\mathbf{k}}-E_{0}}{2}
  \sqrt{1-\frac{(2{\Xi}_{\tau,\mathbf{0}})^2}{\left(2{\Xi}_{\tau,\mathbf{0}}
  -E_{0}\right)^2}}\frac{\Psi_{0,\mathbf{k}}}{\Psi_{0,\mathbf{0}}}.\hskip3ex
\end{eqnarray}
By using $2{\Xi}_{\tau,\mathbf{k}} = \tilde{D}+{k^2}/({2\mu_{\text{X}}})$ and
$E_{0}=\tilde{D}-E_{\text{X}}$, and by assuming the ground-state exciton wavefunction
being given by the ground-state wavefunction of two-dimensional hydrogen atoms,
$\Psi_{0,\mathbf{k}} \simeq
{2\sqrt{2\pi}\mathcal{Z}^2}/{\left(k^2+\mathcal{Z}^2\right)^{3/2}}$, the gap-order
parameter can be given approximately by
\begin{eqnarray}
  \Delta_{\tau,\mathbf{k}}
  &\simeq&
  \pm\theta(E_{\text{X}}-\tilde{D})
  \frac{\mathcal{Z}^3\left[E_{\text{X}}+{k^2}/({2\mu_{\text{X}}})\right]}
  {2\left(k^2+\mathcal{Z}^2\right)^{3/2}}
  \sqrt{1-\frac{\tilde{D}^2}{E_{\text{X}}^2}},\n
  \label{order_parameter}
\end{eqnarray}
with $\theta(x)$ being a step function. The variational coefficient $C_{I}$ for $I>0$ and
the higher-order corrections of the gap order parameter can be calculated by using the
Newton iterative method, and it can be shown that the iteration is converged. Since a
nontrivial solution of the gap equation exists, the sufficient condition of exciton
instability $\tilde{D}<E_{\text{X}}$ is shown.

The numerical solution of the gap equation can be obtained by using these formulations and
the Newton iterative method. The exciton wavefunction $\Psi_{I,\mathbf{k}}$ is given by
$\tilde{\Psi}_{I}(\mathbf{k})$ from Eq.~(\ref{exciton_eq0}) in
Sec.~\ref{sec:exciton_binding_energy}. The initial condition of the gap order parameter is
given by Eq.~(\ref{order_parameter}). The convergence of the iteration can be reached
efficiently.



\end{document}